\documentclass[twocolumn, twocolappendix]{aastex631}
\usepackage{amsmath}
\usepackage{mathtools}
\usepackage{xspace}

\DeclarePairedDelimiter\ceil{\lceil}{\rceil}

\newcommand{\themis}{\textsc{Themis}\xspace}
\def\ehtim{\texttt{eht-imaging}\xspace}
\def\alinet{\texttt{ALINet}\xspace}

\begin{document}
\submitjournal{The Astrophysical Journal}
\title{Validation and Calibration of Semi-Analytical Models for the Event Horizon Telescope Observations of Sagittarius A*}

\shorttitle{Calibrating Semi-Analytical Models for EHT}

\author[0009-0003-4620-8448]{Ali SaraerToosi}
\affiliation{Perimeter Institute for Theoretical Physics, 31 Caroline Street North, Waterloo, ON, N2L 2Y5, Canada}
\affiliation{Department of Physics and Astronomy, University of Waterloo, 200 University Avenue West, Waterloo, ON, N2L 3G1, Canada}
\affiliation{Department of Computer Science, University of Toronto, 40 St. George St., Toronto, ON, M5S 2E4, Canada}
\author[0000-0002-3351-760X]{Avery E. Broderick}
\affiliation{Perimeter Institute for Theoretical Physics, 31 Caroline Street North, Waterloo, ON, N2L 2Y5, Canada}
\affiliation{Department of Physics and Astronomy, University of Waterloo, 200 University Avenue West, Waterloo, ON, N2L 3G1, Canada}
\affiliation{Waterloo Centre for Astrophysics, University of Waterloo, Waterloo, ON N2L 3G1 Canada}

\begin{abstract}
    The Event Horizon Telescope (EHT) enables the exploration of black hole accretion flows at event-horizon scales. Fitting ray-traced physical models to EHT observations requires the generation of synthetic images, a task that is computationally demanding. This study leverages \alinet, a generative machine learning model, to efficiently produce radiatively inefficient accretion flow (RIAF) images as a function of the specified physical parameters. \alinet has previously been shown to be able to interpolate black hole images and their associated physical parameters after training on a computationally tractable set of library images. We utilize this model to estimate the uncertainty introduced by a number of anticipated unmodeled physical effects, including interstellar scattering and intrinsic source variability.  We then use this to calibrate physical parameter estimates and their associated uncertainties from RIAF model fits to mock EHT data via a library of general relativistic magnetohydrodynamics models.
\end{abstract}
\section{Introduction}\label{sec:intro}
The Event Horizon Telescope (EHT) is a very long baseline interferometer \citep[VLBI][]{2019BAAS...51g.256D, Akiyama_2019} that was created to facilitate studying the astrophysics and gravity of a black hole within the resolution of the event horizon \citep{Broderick_2009, Broderick_2011, Broderick_2014, Broderick_2016, Volkel:2019muj, Falcke_2000}. 
Thus far, the EHT has successfully imaged Sagittarius A* (Sgr A*; \citet{sgrA_paper1}) and Messier 87* (M87*;\citet{1906.11238}), revealing or verifying information about the metric induced by the black hole \citep{Johannsen_2016, Broderick_2014, m87_paper6, Kocherlakota_2020, sgrA_paper6, salehi2023photon, Broderick_Salehi}, spin of the black hole \citep{0508386, Broderick_Loeb_2006a, Broderick_2009, Broderick_2011, Broderick_2016}, its mass \citep{1906.11238, sgrA_paper4}, and the dynamics of the accretion processes surrounding the black hole \citep{Broderick_Loeb_2005, Broderick_Loeb_2006a, Tiede_2020, Gerogiev_2022, Broderick_2016_dynamics, Broderick_2022, Ni_2022}.

The EHT produces images from visibilities, which are mathematically connected to the sky image through a Fourier transform. Ideally, a complete knowledge of the visibilities would provide a complete and accurate characterization of the image of the astronomical object \citep[Chapter~2]{Thompson2017}. However, in practice, the visibilities cover only a limited range of spatial frequencies, which are determined by the relative positions of EHT stations on the sky. To overcome this challenge, forward modeling techniques are used to bypass direct image comparison, as well as to account for a different uncertainty systematics; instead, these methods directly compare observed visibilities with those predicted by theoretical models \citep{m87_paper6, sgrA_paper4, Broderick_2009}. This approach is advantageous because it uses well-defined, nearly Gaussian uncertainties in the visibilities, which are independent and essential for constructing accurate likelihoods. 

The choice of model determines how to handle gaps in the visibility data, 
exploring in a well-defined manner the ambiguities these gaps might introduce. 
By using physically based models, that is, those that simulate orbiting plasma and gravitational lensing in the Kerr spacetime, it is possible to directly infer the physical properties of black holes and the characteristics of the surrounding accretion flows. 
Because of this, direct physical modeling has been natively incorporated into \themis since its inception \citep[see Sections 4.4, 8.3, and 9.2 of][]{Themis}.  A significant challenge posed by the native ray-tracing and radiative transfer modules of \themis is their substantial cost, often requiring many seconds per image evaluation.

It has been empirically shown that Sgr A* is radiatively inefficient.  The meager luminosity is significantly sub-Eddington. Moreover, when compared to the accretion rates inferred from Faraday rotation studies and recent EHT modeling, Sgr A* must have a radiative efficiency less than 0.1\%,  well below the standard 10\% typical of efficiently emitting accreting supermassive black holes \citep{Agol_2000, Quataert_2000, Marrone_2007, Yuan_Narayan_2014, sgrA_paper5}.
Radiatively inefficient accretion flow (RIAF) models, therefore, are
well motivated for the interpretation of Sgr A* and its environment physical properties.

RIAF models are designed to explain the low radiative efficiency observed in highly sub-Eddington sources, such as those targeted by the EHT at the event horizon scale, most importantly Sgr A*. Observations over the past two decades support the existence of RIAFs \citep{narayan_riaf_review, Yuan_Narayan_2014, Broderick_2009, Broderick_Loeb_2006a, narayan_2008}. Such flows are designed to explain systems with accretion rates well below the Eddington rate, i.e., $\dot{M} < 0.01 \dot{M}_{Edd} = 0.2 (M / 10^9 M_{\odot}) M_{\odot} \text{yr}^{-1}$ by taking into account the weak Coulomb coupling between the electrons and the much more massive ions, where the former radiate efficiently and the latter do not \citep{narayan_riaf_review, Yuan_Narayan_2014, Narayan_1998}. 

RIAF models typically feature a thick, nearly virialized disk structure. Based on semi-analytic stationary state models, \citet{BL06} and \citet{Broderick_2011} developed simplified models that are well suited for millimeter-wavelength Very Long Baseline Interferometry (mm-VLBI) imaging, where plasma density, temperature, and magnetic fields are described using radial power laws and vertical Gaussian profiles \citep{Broderick_2009, Broderick_2011, Broderick_2016}. A more generalized set of RIAF models, where parameters such as power-law indices, vertical scale height, and orbital velocity are considered adjustable, was introduced by \citet{Broderick_2020, Pu_2018}. The creation of model images involves ray tracing, a process in which photon paths are traced backward in time from each pixel on a distant screen, and the equations of relativistic radiative transport are solved along these paths \citep{BroderickBlandford2003, BroderickBlandford2004, ipol_paper, Gold_2020, PhysRevD.94.084025}.

To extract meaningful information about source parameters from EHT data, a Bayesian parameter estimation framework is employed \citep{Broderick_2020}. This method samples the flikelihood of the model to derive posterior distributions for the model parameters directly from the data. In the case of RIAFs, these parameters, such as the black hole’s spin, mass, and inclination angle, are crucial for understanding the system’s physical nature. However, this process is computationally intensive, as each step requires solving the equations of polarized radiative transfer to generate the corresponding images. Even with semi-analytical models like RIAFs, which avoid the need for extensive simulations to produce plasma distributions \citep{Broderick_Loeb_2006a, Pu_2018}, generating the approximately one billion images required for a single EHT data analysis is a significant challenge. Repeating this process hundreds of times to assess model bias and systematic errors is nearly impossible, even with the use of GPU acceleration and advanced parallel computing techniques \citep[see, e.g.,][]{Broderick_2020, Tiede_2020}. However, because variations in the RIAF model parameters lead to smooth and continuous changes in the resulting images, interpolation methods can be employed to reduce the computational burden. Given the complex and nonlinear relationship between model parameters and the images they generate, a nonlinear interpolation approach is necessary to make this process more efficient.

In a prior publication, \citet{SaraerToosi_2024}, a machine learning method was presented, \alinet, which is capable of generating simulation images given its physical parameters. Here, we use an \alinet trained on extensive raster of RIAF images. This machine learning model replaces the computationally expensive radiative transfer equations that need to be calculated each time an image is to be generated, thereby reducing the time it takes to generate a single image from a few minutes to a few milliseconds. However, \alinet only generates images given an astrophysical models parameters. The ultimate goal is to constrain the physical parameters of data, synthetic and observational.

In order to find and constrain the physical parameters of simulation or observational data, we use \themis developed by \cite{Themis}. \themis is a computationally efficient software platform that, through the utilization of parallel tempering Monte Carlo sampling of a geometric or physical model, constrains the values associated with that model. For our purposes, we use the RIAF-trained \alinet to generate images within \themis. Through Monte Carlo sampling of the \alinet model, we constrain the physical values of the data. This paper is the second in a series of papers, the first of which being \citet{SaraerToosi_2024}; the same RIAF simulations used in the first paper are utilized here to verify the performance of the parameter estimation framework of the \alinet within \themis.

RIAFs are relatively computationally inexpensive to simulate. 
However, large libraries of general relativistic magnetohydrodynamics (GRMHD) simulations that explore the dynamical evolution of the accretion flow are now available \citealt{sgrA_paper5,Porth_2019, Athena, BHAC_Porth_2017, BHAC_Ripperda_2019a, BHAC_Ripperda_2019b, Cosmos++_Anninos_2005, Cosmos++_Fragile_2012, Cosmos++_Fragile_2014, ECHO_Del_Zanna_2007, ECHO_Del_Zanna_2011, ECHO_Del_Zanna_2012, ECHO_Del_Zanna_2018, H-AMR_Gammie_2003, H-AMR_Noble_2006, H-AMR_Liska_2017, H-AMR_Chatterjee_2019, iharm3D_Noble_2009, IllinoisGRMHD_Etienne_2015}.  Importantly, these relax assumptions made in the semi-analytic RIAF models by self-consistently modeling the magnetic interactions responsible for angular momentum transfer in the accretion flow. 
Nevertheless, the GRMHD prescription incorporates a number of strong assumptions that are likely to be violated in Sgr A*.

First, the hydrodynamic approximation assumes a locally isotropized distribution function due to plasma instabilities, though thermalization is typically prescribed through an equation of state. However, in low-density regions near black holes, the mean free paths of particles can become comparable to the system size, leading to collisionless plasma behavior that is not accurately captured by fluid models.

Second, the vast majority of GRMHD simulations do not explicitly simulate the radiating electrons, which are subsequently ``painted'' onto the simulation post-hoc, typically ignoring their acceleration and radiative cooling \citep[see, e.g.,][]{Porth_2019}.  Even when two-temperature flows are simulated, the acceleration of nonthermal electrons are not incorporated, despite their crucial role in producing much of the observed emission.  As a result, GRMHD simulations typically do not replicate flaring behaviors seen in systems such as Sgr A* \citep{Ripperda_2022, Ripperda_2020}, and those that approximate flaring rely heavily on interpretive frameworks.  
This limitation arises partly from the inability to resolve the dissipative scales where the MHD approximation breaks down.  That is, MHD models fail to capture kinetic-scale effects such as magnetic reconnection, turbulence-driven stochastic acceleration, and relativistic shocks mechanisms that are crucial for energizing electrons to the Lorentz factors needed to produce the observed synchrotron emission. These acceleration processes require a kinetic or at least a two-temperature fluid treatment, which standard GRMHD simulations do not include\citep{Broderick_2015}.

Other limitations of GRMHD simulations include the widespread use of the ``fast light'' approximation, restricted computational domains, the monotonic decrease in plasma density as it accretes onto the black hole, and extremely limited imposition of spectral constraints beyond the total flux at 1.3~mm.
\citep{sgrA_paper4}.  All of these restrictions are naturally addressed in the semi-analytic RIAF models \citep{Broderick_2016}, at the not-insignificant expense of failing to self-consistently capture the dynamical nature of the accretion flow.

Because neither the computationally inexpensive semi-analytic RIAF models nor the physically limited GRMHD simulations can provide a fully self-consistent picture of the Sgr A* accretion flow, it is useful to explore the robustness of parameter inferences to variations in the underlying modeling methodology.  That is, given simulated data generated from a GRMHD simulation, can the RIAF analyses enabled by \alinet recover the correct black hole parameter estimates?  By addressing this question here, we quantify the robustness of the semi-analytic RIAF model to modifications in the underlying emission morphology.

In this paper, we first run a series of tests to verify the fidelity of a RIAF-trained \alinet model within \themis against mock data generated from RIAF simulations in \autoref{sec: self consistency tests}. In \autoref{subsec: scatt RIAF test}, we assess its performance against RIAFs that have been subject to diffractive and refractive scattering. In \autoref{sec: model misspecification tests}, we test the performance of \themis and an \alinet that has been trained on the stationary RIAF image library against dynamic GRMHD simulations. Afterward, In \autoref{sec: Calibration Set}, we quantify how RIAFs can constrain the physical parameters of a GRMHD simulation, such as spin and mass. Finally, we finish by discussing how these uncertainties should be taken into account when a semi-analytical model such as RIAF is used in fitting to observational data subject to scattering.

\section{Astrophysical Models}\label{sec: RIAF Simulations}
\subsection{RIAF Models}
RIAF models offer a framework for understanding the process by which matter accretes onto black holes in scenarios where the system exhibits low luminosity or is in a relatively quiescent state. Unlike more efficient accretion modes, RIAFs convert only a small fraction of the gravitational potential energy into radiation, with most of this energy being advected towards the black hole's event horizon rather than being emitted as light \citep[for a detailed review, see][]{narayan_riaf_review}. The reduced efficiency of radiative cooling in RIAFs leads the ions in the gas to heat up to nearly virial temperatures, reaching as high as $10^{12}~{\rm K}$ near the black hole \citep{Yuan_2003, Broderick_seven_years}. When the accretion disk lacks a dominant large-scale magnetic field, this high-pressure gas forms a thick, geometrically extended disk, commonly referred to as the standard and normal evolution (SANE) state. However, if the black hole accretes a substantial amount of magnetic flux, the resulting magnetic pressure can compress these hot disks into a more compact structure, known as a magnetically arrested disk (MAD) state.

RIAFs primarily produce radio emission through synchrotron radiation. In the extreme conditions of RIAFs, the high gas temperatures mean that thermal synchrotron radiation is a significant contributor, with the electron temperature $kT_e$ far exceeding the electron rest mass energy $m_e c^2$. However, as with many astronomical sources of synchrotron radiation, a large portion of the observed emission, particularly from objects observed by the Event Horizon Telescope (EHT), likely originates from non-thermal electrons. These electrons can be accelerated by various energetic processes, such as magnetic reconnection, shocks, or other types of violent dissipation events \citep{Yuan_2003, Yuan_2009}. To accurately model the emission from RIAFs, it is therefore necessary to account for both the thermal and nonthermal components of the electron population.

The RIAF model applied in our study is based on a parameterized framework that was initially developed to interpret early millimeter wavelength very long baseline interferometry (mm-VLBI) observations \citep{0508386, Broderick_2009, Broderick_2011, Broderick_2016} and has since been refined and expanded in subsequent work \citep{Pu_2018, Broderick_2020}. This model, inspired by \citet{Yuan_2003}, incorporates radial power-law spatial distributions for both thermal and nonthermal plasma components, with a toroidal magnetic field strength set at a specific fraction of
equipartition (normally for plasma $\beta=10$, where $\beta$ is the ratio of gas pressure to magnetic field pressure in the disk).
 
The model assumes that the accretion flow orbits azimuthally outside the innermost stable circular orbit, with an angular velocity that is a fixed proportion of the relativistic Keplerian value and then plunges inward 
on constant angular momentum orbits. 
A Gaussian vertical structure is assumed for the disk, with the scale height being a set fraction of the cylindrical radius, allowing the model to simulate either the SANE or MAD states. For our analysis, we focus on five critical parameters: the black hole spin $a$, the cosine of the inclination angle $\mu$, the height-to-radius ratio of the disk $H/R$, the normalization of the number density of nonthermal electrons $n_{nth}$, and the sub-Keplerian fraction $\kappa$. The remaining parameters are determined as functions of $a$ and $\mu$, following the approach outlined in \citet{Broderick_2016}.

\subsection{Image Models}
Images of the physical model of RIAFs outlined in the previous section may be generated using standard ray tracing and radiative transfer techniques. We assume that the primary emission mechanism is synchrotron emission arising from both the thermal and nonthermal electron populations \citep{BroderickBlandford2003,BroderickBlandford2004,Broderick_2016}.  While directly generating images from the underlying is possible within \themis, here we make use of the \alinet models described in \citet{SaraerToosi_2024}.  That is, the images we use are generated as the output of a neural network whose learned weights implicitly encode the underlying physical relationships between image features and physical parameters, like spin $a$ and mass $M$ \citep[see][for details]{SaraerToosi_2024}. As previously noted, \alinet is many orders of magnitude faster than direct image generation, even accounting for the training (which must be done only once and on a set of images far smaller than that typically necessary for parameter estimation).

\section{Data Generation and Analysis}
\subsection{Data Generation}
Given an image, either from a RIAF model with specified parameters or other simulations, we generate simulated data sets for validation, calibration, and testing \alinet and \themis.  We do this with the \ehtim package\footnote{\texttt{https://github.com/achael/eht-imaging}}, which incorporates realistic noise and station-based corruptions \citep{Chael_2016, Chael_2018, Chael_2022,ehtim:23}.  We use the same baseline coverage as the 6, 7, 10 and 11 April 2017 EHT Sgr~A* observations \citep{sgrA_paper2}.
For tests where the underlying image is scattered, this is done using the scattering screen model in \citet{Johnson_2018}, with the parameters in \citet{Issaoun_2021}, as implemented in the \texttt{StochasticOptics} package within \ehtim.
Simulated data is subsequently preprocessed as described in \citet{sgrA_paper4}, and prepared for analysis with \themis using the ThemisPy package\footnote{\texttt{https://github.com/aeb/ThemisPy}}.

\subsection{Analysis}
In our analysis, we utilize \themis for Bayesian parameter estimation, employing a Markov Chain Monte Carlo (MCMC) approach enhanced with parallel-tempered sampling algorithms. This technique effectively explores complex multimodal posterior distributions, which are anticipated in our parameter space \citep[see, e.g.,][]{Broderick_2016}.

Due to the small number of RIAF model parameters, we make use of the Automated Factor Slice Sampling (AFSS) exploration kernel.  AFSS is efficient when dealing with modest dimensional parameters spaces and complex posterior probability distributions.  By adaptively determining the sampling steps, AFSS enhances the exploration of the parameter space, leading to more accurate and reliable parameter estimation \citep{afss}.
We employ parallel tempering to globally explore the model parameter space, efficiently identifying and characterizing multimodal posteriors.  Communication among tempering levels is facilitated via the deterministic even-odd swap (DEO) scheme, which efficiently traverses the temperature ladder without the limits imposed by diffusion schemes \citep{DEO:2019}.

In addition \themis provides a number of tools for addressing the various systematic uncertainties present in EHT data sets.  Complex station gains, e.g., due to atmospheric absorption and phase delays, are marginalized over via the Laplace approximation, variability and refractive scattering are mitigated via additional model-dependent noise budgets, and diffractive scattering removed via deconvolution with the measured scattering kernel, all as described in \citet{sgrA_paper3} and \citet{Broderick_2022}.

We initialize each analysis with all physical parameters set to zero and allow the sampling to proceed until convergence,
assessed using established criteria as described in \citet{Themis} and \citet{m87_paper6,sgrA_paper4}.

\section{Self Consistency Tests}\label{sec: self consistency tests}
Our first set of tests involves testing the \alinet trained on the RIAF data set against the output of the \alinet model itself and new RIAF images that have been generated by solving radiative transfer equations. The former test is to validate that our method is self-consistent, i.e., that given an \alinet-generated image, we are able to recover the physical parameters within (typically narrow) posteriors through our fitting procedure.\footnote{Here and elsewhere, we note that by recover, we mean that the bulk of the posteriors, which we observe to be typically very narrow, covers the truth values.} This test confirms that there are no unanticipated degeneracies within the range of acceptable values for the physical parameters,
i.e., there is always one and only one
image for a specific set of physical parameters. The latter test, namely testing the \alinet model on a set of RIAF images that the model has not seen before (the testing dataset), confirms that the implementation of the \alinet model within \themis is indeed capable of recovering the physical parameters of images it has not seen before through interpolation. 

We use different visualizations to show the results; we use triangle plots for one test case for each of the tests, while showing the corresponding truth and best fit images of that test. We also show fitted parameter distributions for three data points for the scattered RIAF test, mean GRMHD test, and variable GRMHD test. We show more fitting results for all of the tests in \autoref{sec: Tests appdx}. 
Furthermore, comparisons of the fit to mock visibility data can be found in \autoref{subsec: Complex visibility residuals appdx}, where residuals are presented for some examples. In all the following sections, we show one instance of that test as a representative.
\begin{figure*}[!ht]
    \centering
    \includegraphics[width=1\textwidth]{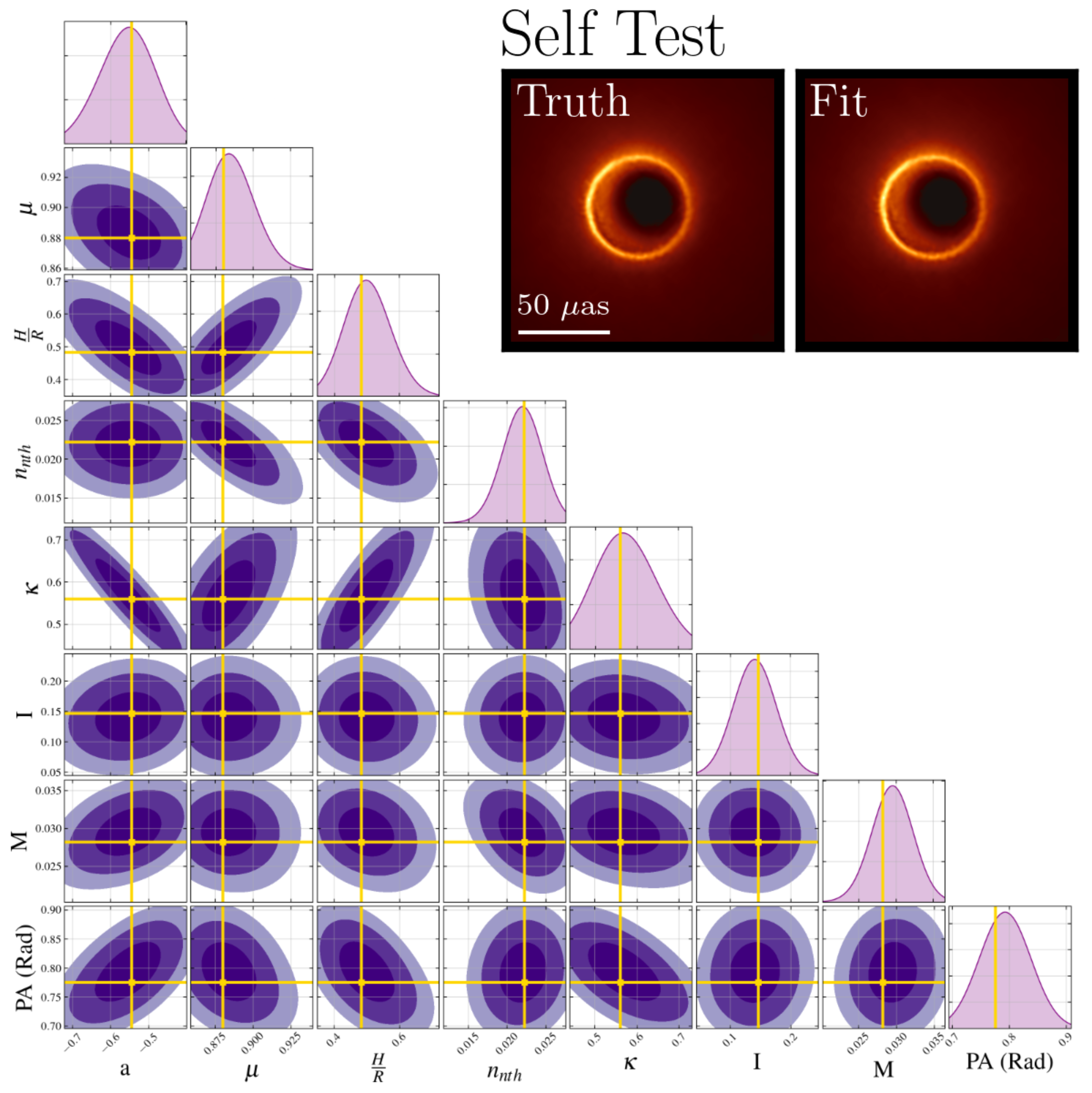}
    \caption{\alinet Self test truth image vs. best fit image from \themis, and joint posteriors for a representative \alinet self test.  Contours show $68.3\%$, $95.4\%$, and $99.7\%$ cumulative probability regions, indicating 1, 2, and 3$\sigma$ confidence levels, respectively. Gold points indicate the truth values.  
        \label{fig: self test}}
\end{figure*}

\subsection{\alinet Self Test}\label{subsec: self test}
The recovered image and the parameters for this image for this test is shown in \autoref{fig: self test}. For more tests, see \autoref{sec: Tests appdx}. Note that the image shown in \autoref{fig: self test} is the fifth image (from the left) of the test figures in \autoref{subsec: RIAF Tests appdx}. We choose the data generated from the same RIAF parameters, that is, the first $5$ parameters out of the total $8$ are the same for all tests shown in the main text, and all correspond to the fifth image of their corresponding plots in the \autoref{subsec: RIAF Tests appdx}. We do not show the results for the parameter fits for more than one case; however, the \themis fitted distributions perfectly recover the truth in all cases. In \autoref{fig: self test}, we see that in the representative example, all parameters are recovered as expected. We also see the same behavior in all the self tests that we perform. The success of the \alinet self test shows that the RIAF-trained \alinet recovers the parameters of the images produced by \alinet within the \themis framework.

\begin{figure*}[!ht]
    \centering
    \includegraphics[width=1\textwidth]{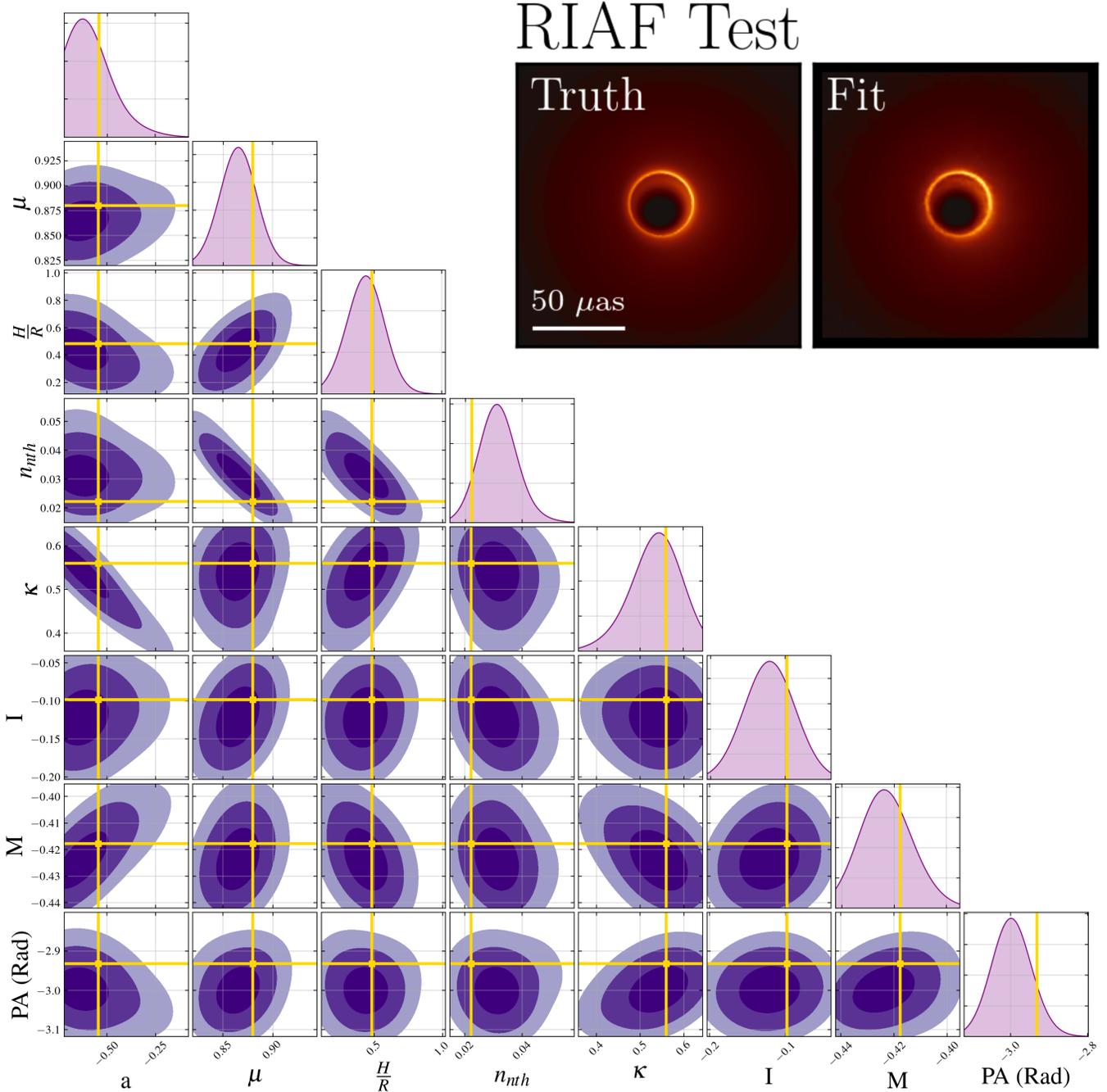}
    \caption{RIAF test Truth image vs. best fit image from \themis, and parameter fits triangle plots. The gold circles in the triangle plot denote the truth values. \label{fig: RIAF test}}
\end{figure*}

\subsection{RIAF Test}\label{subsec: RIAF test}
 The previous test shows that the underlying physical parameters of RIAF images that \alinet has learned is recoverable using \alinet within \themis. Here, we demonstrate how \alinet within \themis can also fit to and recover the parameters of RIAF images on which \alinet has not been trained within its posteriors. The results of a RIAF test are shown in \autoref{fig: RIAF test}. We perform this test with the same set of physical parameters as for the \alinet self test. For more RIAF tests, see \autoref{fig: RIAF tests fit_and_truth}. We do not show the results for the parameter fits for more than one case; however, the \themis fitted distributions satisfactorily recover the truth. The success of the RIAF test shows that the RIAF-trained \alinet within \themis is able to recover the parameters of all RIAF images generated via simulation.
\begin{figure*}[!ht]
    \includegraphics[width=1\textwidth]{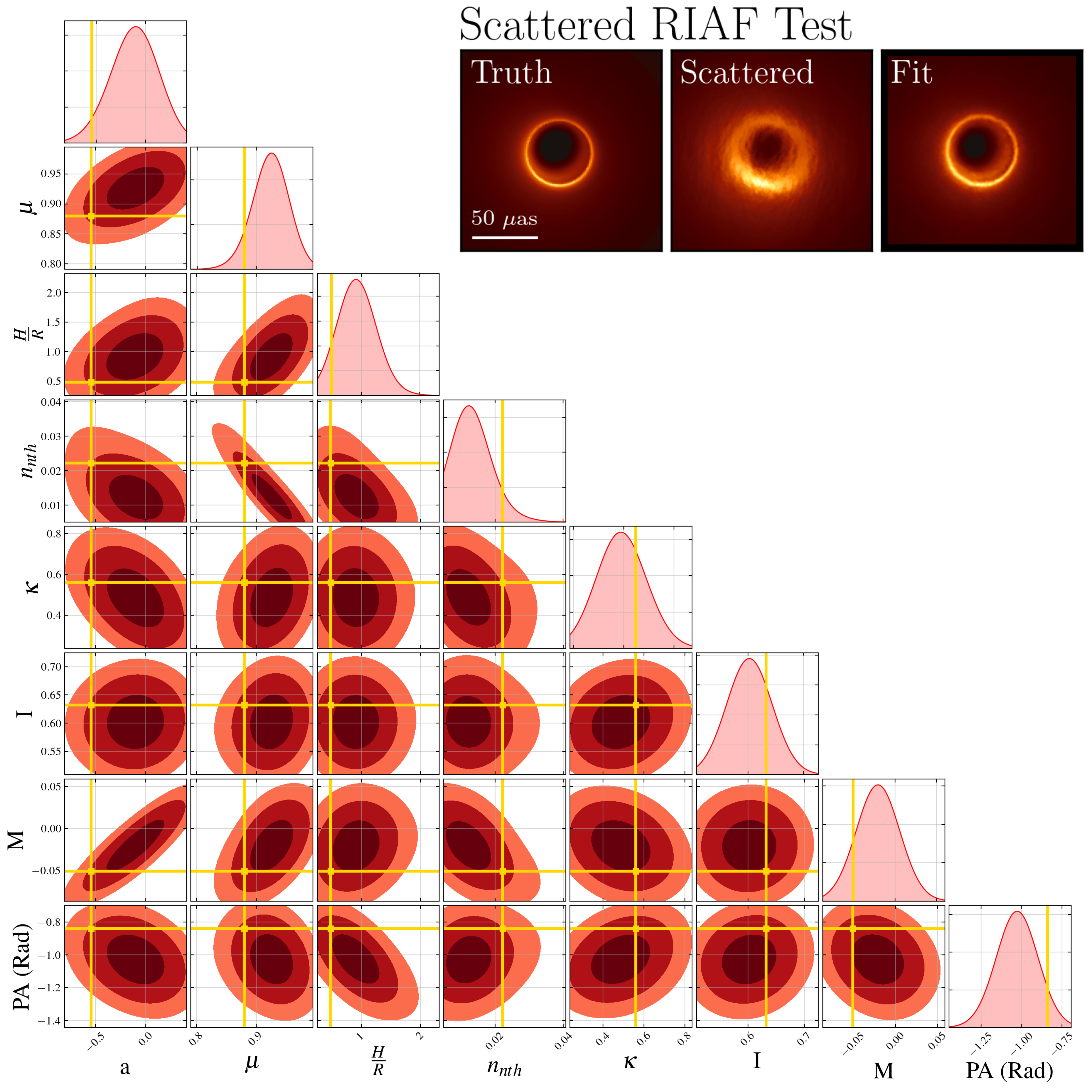}
    \caption{Scattered RIAF test Truth image, scattered image to which fitting is done, best fit image from \themis, and parameter fits triangle plots. The gold circles in the triangle denote the truth values. The threshold noise used is $7$ mJy.\label{fig: RIAFscatt}}
\end{figure*}
\begin{figure*}[h!]
    \centering
    \includegraphics[width=0.8\linewidth]{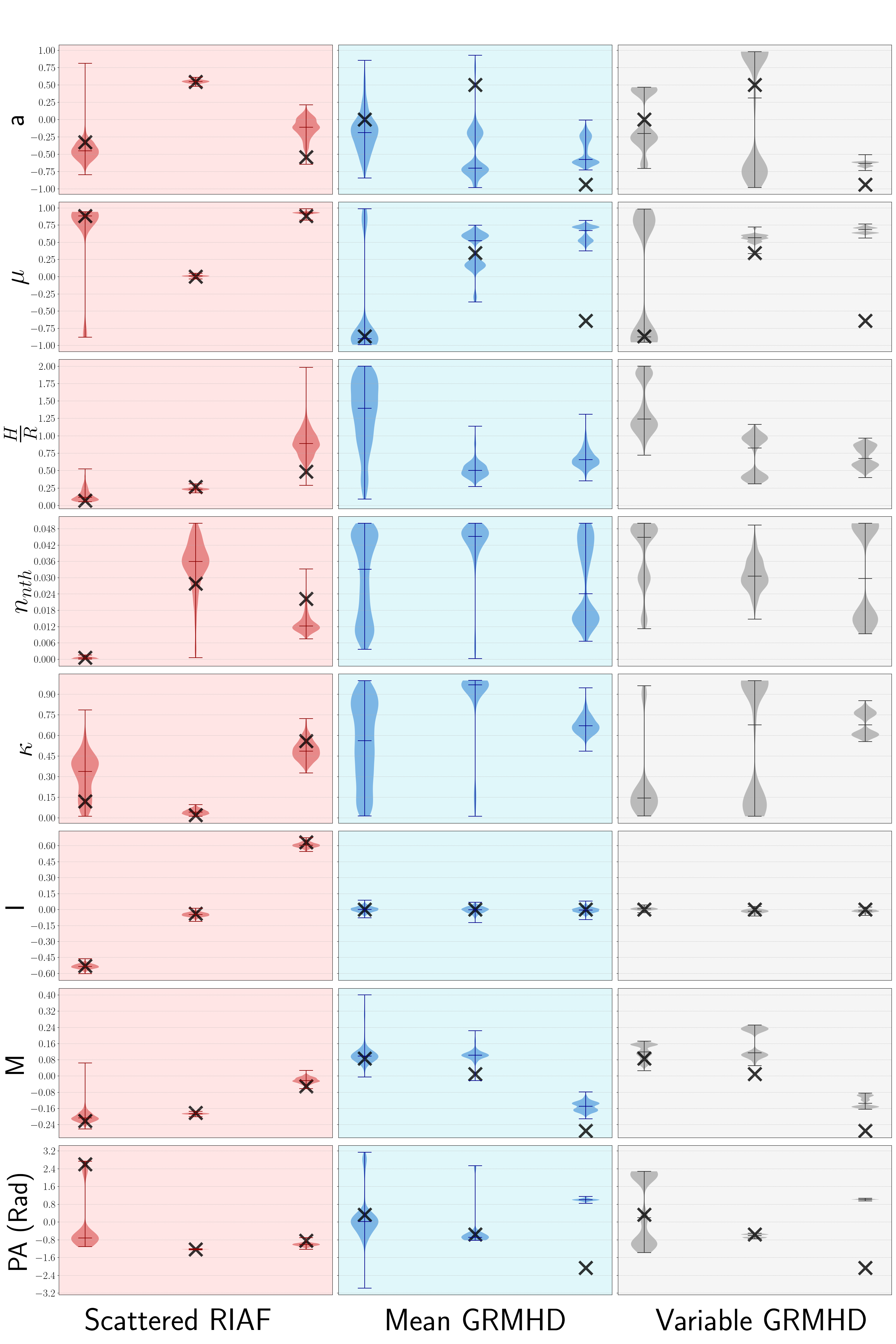}
    \caption{Violin plots for three test cases for scattered RIAF test (on the left), mean GRMHD (middle), and variable GRMHD (right). The X shows the truth value. The violins show the resulting fitted distributions from \themis. The bars in each distribution show minimum, maximum, and the median values of the distribution. In each column, the first two from the left are MADs, and the third is a SANE.}
    \label{fig: scatt,grmhd violins}
\end{figure*}

\subsection{Scattered RIAF Test}\label{subsec: scatt RIAF test}
The effects of scattering on observation are significant. Scattering introduces additional small-scale structures which blurs out intrinsic features. Here, we study this additional uncertainty through fitting the RIAF-trained \alinet model to data generated from a set of scattered images.
For these scattered tests, we choose RIAF images, scatter the images using the scattering screen model in \citet{Johnson_2018}, using the parameters in \citet{Issaoun_2021}, as implemented in the \texttt{stochastic\_optics} package within \ehtim and create simulated data using 2017 April 6 and 7 EHT coverage.

We again assess how well the underlying physical parameters of the truth image can be recovered.
The fitting process is modified to account for the impact of scattering in two ways: 
1. The introduction of diffractive scattering kernel that blurs the sampled image, and 2. the inclusion of an additional uncertainty in quadrature with the stochastic (thermal) and systematic error terms to address the additional refractive substructures (see \citet{sgrA_paper4} for details). Scattering mitigation substantially broadens the resulting posteriors, which was expected given that scattering obscures small-scale lensing features.

We fit to a number of scattered RIAF images with
the \alinet model in \themis.  Good fits are found generally, residual plots are presented in \autoref{sec: Tests appdx}.

Two-parameter marginalized posteriors for a representative case are shown in \autoref{fig: RIAFscatt}.  Additional examples may be found in \autoref{sec: Tests appdx}.

In the case shown in \autoref{fig: RIAFscatt}, the test demonstrates the capacity of the ALINet model to successfully recover the physical parameters of the underlying image in spite the additional complications imposed by scattering.  However, unlike the previous self-tests, this success is not universal, and for some rare cases scattering can result in multimodal and biased posteriors.  This disparity is a direct consequence of the imperfect match between ALINet model and the scattered truth image due to the presence of refractive scattering in the latter.  We address how we determine and apply an additional systematic error budget in \autoref{sec: Calibration Set} and only note here that it is small (see \autoref{table: Calibration set combined}).  These systematic errors should be incorporated before any interpretation of RIAF parameters are made for Sgr A*.

\section{Robustness to Model Misspecification Tests}\label{sec: model misspecification tests}
\subsection{GRMHD Simulations}\label{sec: GRMHD description}
So far, we have successfully shown with a number of mock data tests that RIAFs can be extracted even after addressing the various expected astrophysical and systematic corruptions of EHT data. However, we now turn to the difficulties unique to Sgr A* arising from its intrahour variability. To do this, we make use of GRMHD simulations, which despite their limitations, have the virtue of self-consistently modeling variability in the accretion flow.

GRMHD simulations evolve magnetized charge-neutral gas on a fixed stationary spacetime, thereby modeling the dynamical evolution of the accretion flow onto a black hole. 
They do this by simultaneously solving the equations of particle number conservation, conservation of energy-momentum, and the source-free evolution equation for the magnetic field, constrained by the no-monopole condition \citep{Porth_2019}.
In practice, numerical resistivity and viscosity are introduced to stabilize the numerical evolution, and facilitate magnetic reconnection and the formation of shocks on the grid scale \citep{Porth_2019, Bart_2019}.
The accretion flow plasma is typically treated as a non-radiative ideal gas in the Kerr metric with spin $a$ as a free parameter. \citep{H-AMR_Gammie_2003, Cosmos++_Anninos_2005, ECHO_Del_Zanna_2007, white_2016, BHAC_Porth_2017}. Motivated by the statistically steady state of the near-horizon flow, which is where most of the radiation is produced \citep{m87_paper5}, for almost all of the models, the initial conditions are constant hydrodynamic equilibrium tori with a poloidal magnetic field \citep{Fishbone_1976} with orbital angular momentum along the black hole spin.\footnote{While other initial conditions are also used, namely stellar wind-fed models \citep{Ressler_stellar_winds} and strongly magnetized torus simulations titled \citep{Liska_non_MAD_torus_2018,Chatterjee_non_MAD_torus_2020}, here we study GRMHDs with initial conditions utilized in \citet{sgrA_paper3, sgrA_paper4}.} The adiabatic index is kept constant for all models. 

Most simulations evolve to more than $3\times10^4\,GM / c^2$.
The late-time behavior of GRMHD simulations is typically divided into two states: magnetically arrested disc (MAD; \citet{Bisnovatyi_MAD_1976, Igumenshchev_MAD_2003, Narayan_MAD_2003, Tchekhovskoy_MAD_2011}) and standard and normal evolution (SANE; \citet{DeVilliers_SANE_2003, H-AMR_Gammie_2003, Narayan_SANE_2012}). This division is characterized by the magnitude of the magnetic flux through the horizon, where the accreted magnetic flux onto the hole in MADs is typically more than $60$ times bigger than SANEs, leading to structural and quantitative differences between the two states. The additional instabilities in MADs, where the accretion is more affected by the black hole magnetosphere, make them more numerically challenging to simulate. 

Synthetic GRMHD images are generated via numerical integration of the radiative transfer equations, where the relevant parameters are density scale (or equivalently plasma accretion rate $\dot M$), cosine of the inclination angle $\mu$, position angle PA, and the model dependent physical parameters spin $a$ and mass $M$. Furthermore, the electron distribution function is assumed to be Maxwell-J\"uttner with a temperature set by an ion-electron temperature ratio $R\equiv T_i / T_e$ that depends on the plasma $\beta$. The ions are assumed to be non-relativistic with adiabatic index of $5 / 3$ and the electrons are relativistic with adiabatic index of $4 / 3$. Here, the same assumptions as made in \citet{m87_paper5} are utilized, where, for regions of $\beta << 1$, i.e., regions where gas pressure is much more prominent than magnetization, $R = T_i / T_e \propto R_{\rm hi}$ has only one free parameter $R_{\rm hi}$ which takes the values within the range $[1, 160]$. Note that as $R_{\rm hi}$ changes from $1$ to $160$, the emission is shifted away from the midplane to the poles. 

In summary, the GRMHD simulations subject to different initial conditions, are parameterized by spin $a$, mass $M$, cosine of the inclination angle $\mu$, position angle PA, flux intensity $I$, and a temperature parameter $R_{\rm hi}$; other parameters are held constant. In RIAFs, we assume that the electron temperature varies spatially with a normalization set by the observed spectral energy distribution (SED), so we do not use \alinet to fit to the temperature. However, we have three extra parameters, namely height-to-radius ratio of the accretion flow $H / R$, density fraction of the non-thermal electrons $n_{nth}$, and the subkeplerian fraction $\kappa$.
In the tests of the robustness of black hole parameter estimation and sensitivity to variability presented in the following sections, we place no constraints on these three parameters as they are not directly defined in the GRMHD simulations; limiting our tests to the overlapping parameters between the two models, that is, we test whether RIAFs could estimate the parameters common to both GRMHD and RIAF models. 

We utilize the dataset created from the GRMHD simulation library in \citet{sgrA_paper5} for our tests. 
Here we fit our \alinet model to the mean GRMHD validation data set of \cite{sgrA_paper4}. Diffractive and refractive scattering are both applied for all the scattered tests. For variable tests, we incorporate variability mitigation as described in \citet{sgrA_paper4, sgrA_paper5}. 

Furthermore, we use two separate \alinet models to do the fitting; one ``optically thick'' that has been trained on RIAF images with a total flux intensity of $2.4$~Jy and one ``optically thin'' with RIAF training images of flux intensity $0.1$~Jy. We are motivated to use these two models since in our fits, we observe that the underlying GRMHD images appear optically thinner than their RIAF counterparts with the same flux intensity. 
While the origin for this effect is not fully clear, a potential origin is the inhomogeneity in the turbulent emitting plasma in the GRMHD simulations.  Due to the nonlinearity in the radiative transfer coefficients, the emission from high-density clumps can produce the observed flux with lower average densities than those required in the smooth RIAF model; the optical depth in the intervening voids is typically negligible.  
This phenomena is well known in other astronomical contexts, e.g., cluster X-ray studies \citep[see, e.g.,][]{xray_studies1, xray_studies2, xray_studies3}.  There a clumping factor is introduced to boost emission relative to the mean absorption, approximately modeling the nonlinear impact of the inhomogeneties.  Here we only attempt to bound the possible impact of the inhomogeneities by exploring models with and without significant optical depth, obtained by lowering the total flux of the images used to construct the training library.

For demonstration purposes, in each section we present one test case in detail. We show many more test cases in \autoref{subsec: GRMHD Tests appdx}. The test case shown in all the sections that involve GRMHD tests are all the third column of \autoref{fig: val GRMHD tests fit_and_truth}.
\begin{figure*}[!ht]
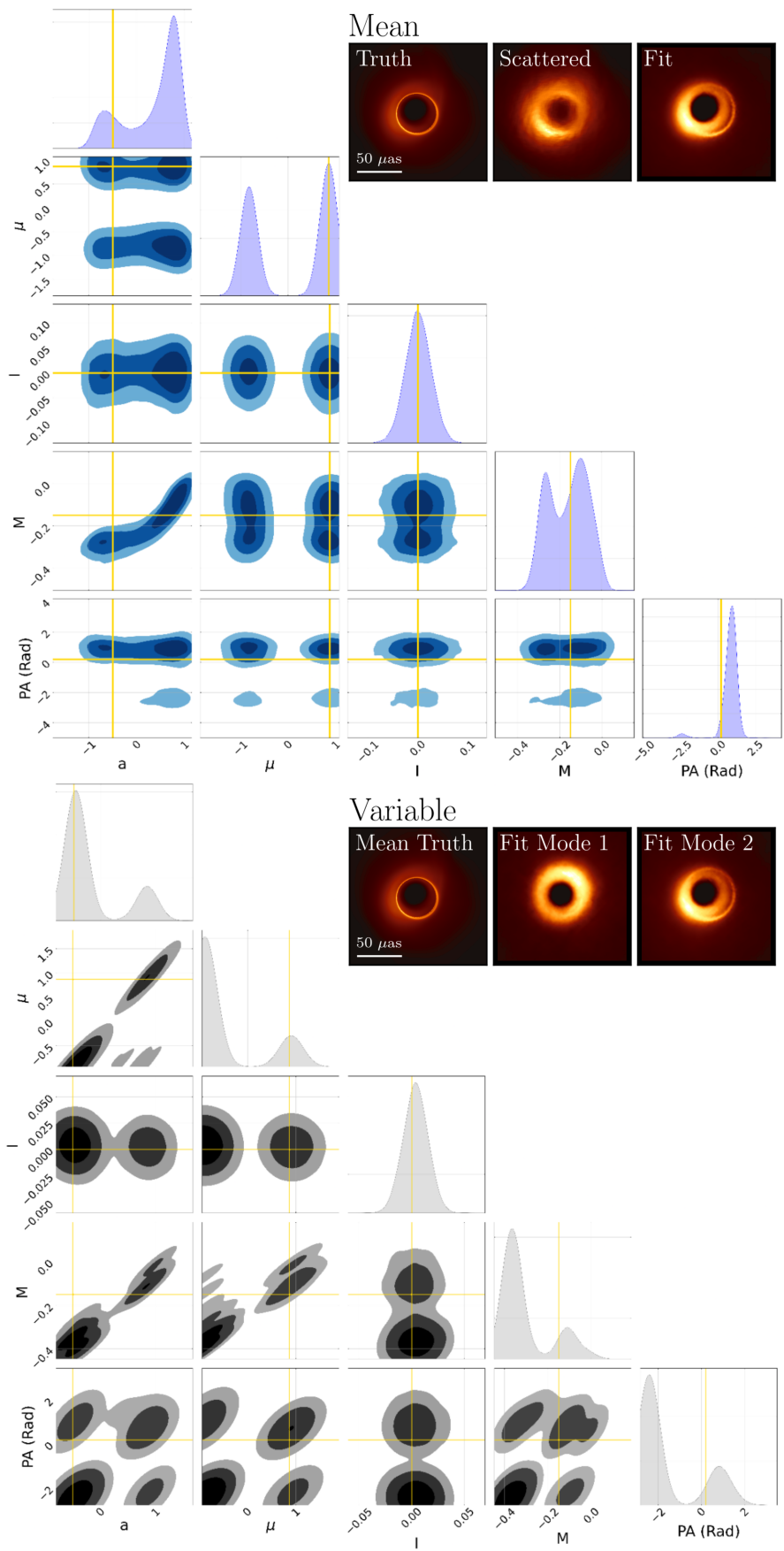

    \fig{GRMHD_triangle_images.png}{0.65\textwidth}{}
    \caption{Scattered GRMHD test including truth image, scattered image to which fitting is done, best fit to mean GRMHD data from \themis (top), truth image and best fit modes (two modes) to variable GRMHD data from \themis (bottom). We used a threshold noise of $7$mJy and $20$mJy for fitting to the mean GRMHD and variable GRMHD case, respectively.\label{fig: mvargrmhd}}
\end{figure*}
Our purpose here is not to present the GRMHD simulations as more physical or complete models of the accretion flow in Sgr A*. Rather, it is to assess the ability of semi-analytical models like RIAFs to robustly extract physical parameters in the face of the anticipated source variability and inevitable misspecification of the model.

\subsection{Scattered Mean GRMHD}\label{subsec: scatt Mean GRMHD}
We begin by fitting optically thick and thin static RIAF images to mean GRMHD images after scattering. The optically thick and optically thin fitting to a representative case is shown in \autoref{fig: mvargrmhd}. For fitting results to the entire validation data set look at \autoref{subsec: GRMHD Tests appdx}. 

A significant contrast with the results from \autoref{sec: self consistency tests} is the appearance of multiple modes in the parameter posteriors.  In some instances, this is unsurprising -- images in which the line-of-sight component of the spin flips sign (i.e., $\mu\rightarrow-\mu$) appear very similar; when only visibility amplitudes are known there is a $\pi$ degeneracy in the PA, implying that the much less well constrained phase information is critical \citep[see, e.g., the extensive discussions in][]{Broderick_2016}. Nevertheless, the multimodality requires extra care when interpreting fitting results from one model to another; the errors due to scattering and introduction of different physics from a new model causes degeneracy in the fitting process. It is common to see that one of the modes of the posteriors looks similar to the true image and has similar physical parameters, while other modes represent conditions that are significantly different.

Note that both of the RIAF-trained \alinet models fail to recover the parameters for some of the MAD cases. However, we can improve the fitting results by noting that the various sources of noise, e.g., scattering and variability, lead to larger noise in the data than the clean-cut RIAF self tests from which we started this work. With bigger floor threshold noise ($20$mJy), i.e., by ignoring the contribution of structures in the images that have a flux intensity of below a certain threshold, the fitting parameter distribution does encompass the truth parameter values.

Furthermore, some of the SANE cases also fail. This is due to the dominance of off-disk emission geometry in majority of the SANE models which is not well defined within a RIAF model, which assumes disk-dominated emission. The thin-disk assumption in the MAD models causes the material to predominantly accrete in the disk plane of the black hole, leading to a similar emission structure to the RIAFs. This similarity in the emission structure leads to better performance when RIAFs are fitted to MADs as compared to SANEs (see figure 4 of \citet{m87_paper5} for more detail).

\subsection{Scattered Variable GRMHD}\label{subsec: Scatt Variable GRMHD}
Finally, we incorporate the key feature of the GRMHD simulations, their variability.  Because we are fitting static images to data taken from dynamic movies, we are attempting only to reconstruct a suitably defined ``average'' image, the meaning of which is complicated by the specific temporal and baseline coverage.  Nevertheless, explicit and quantitative comparisons may be made using the black hole parameter estimates.

As with the mean GRMHD images, we fit the optically thick and thin RIAF models, bounding the impact of the turbulent inhomogeneities in the simulations.  We use the scattered GRMHD movies from the validation set in \citet{sgrA_paper4}, from which an illustrative example is shown in \autoref{fig: mvargrmhd}.  For the results for the entire validation data set, look at \autoref{fig: val GRMHD tests fit_and_truth}.

Like the previous tests, the posteriors are multimodal, driven in part by the additional uncertainty added as part of the variability mitigation scheme \citep{Broderick_2022}.  In \autoref{fig: mvargrmhd} we show the images associated with two of the modes, corresponding to the $(\mu,{\rm PA})=(X,Y)$ and $(A,B)$ posterior peaks.  The most significant difference is the position angle of the brightness maximum, though other significant difference are present (e.g, the mass of the black hole).

Similar trends to the mean-GRMHD tests were found: MADs are better represented by the RIAF model than SANEs, and thus the parameter extraction can be done with higher fidelity for MADs.

\begin{table}[p]
\vspace{5cm}
\begin{rotatetable}
    \begin{deluxetable*}{lcc|cc|cc|cc|cc|cr}
\tabletypesize{\scriptsize}
\tablewidth{0pt}
\tablecaption{Model misspecification and scattering uncertainties on the variable GRMHD calibration set for thick and thin RIAF models together. The number in the parentheses in front of each dataset category denotes the number of samples that were used to quantify the uncertainties.\label{table: Calibration set combined}}
\tablehead{
 \colhead{Dataset} & \multicolumn{2}{c}{Flux $I$} & \multicolumn{2}{c}{Mass $M$} & \multicolumn{2}{c}{Spin magnitude $|a|$} & \multicolumn{2}{c}{Spin axis $z_a$} & \multicolumn{2}{c}{Spin value $a$} & \multicolumn{2}{c}{Normalized spin vector $n_a$}\\
  \colhead{$(\times 10^{-2})$} & \colhead{$\Delta_I$} & \colhead{$\sigma_I$} & \colhead{$\Delta_M$} & \colhead{$\sigma_M$} & \colhead{$\Delta_a$} & \colhead{$\sigma_a$} & \colhead{$\Delta_{z_a}$} & \colhead{$\sigma_{z_a}$} & \colhead{$\Delta_{a}$} & \colhead{$\sigma_{a}$} & \colhead{$\Delta_{n_a}$} & \colhead{$\sigma_{n_a}$}\\
  \colhead{$(\times 10^{-2})$} & \colhead{$2\sigma$/$3\sigma$} & \colhead{$2\sigma$/$3\sigma$} & \colhead{$2\sigma$/$3\sigma$} & \colhead{$2\sigma$/$3\sigma$} & \colhead{$2\sigma$/$3\sigma$} & \colhead{$2\sigma$/$3\sigma$} & \colhead{$2\sigma$/$3\sigma$} & \colhead{$2\sigma$/$3\sigma$} & \colhead{$2\sigma$/$3\sigma$} & \colhead{$2\sigma$/$3\sigma$} & \colhead{$2\sigma$/$3\sigma$} & \colhead{$2\sigma$/$3\sigma$}
}
\startdata 
 \multicolumn{13}{l}{\textbf{Thick + Thin RIAF}}\\
 \hline 
 Scattered RIAF (10) & -0.23/ & 0.4/ & 0.90/ & 0.0/ & 9.0/ & 9.0/ & -0.25/ & 0.0/ & -8.5/ & 10/ & 0.6/ & 0.7/\\
 Entire calibration set (90) & -0.74/1.9 & 0.0/0.80 & -12/-15 & 20/24 & -4.0/0.86 & 39/34 & -28/-43 & 17/15 & -28/-27 & 59/61 & -13/-29 & 21/21\\
 All yellow and green (35) & 1.3/1.5 & 0.0/0.52 & -1.1/-15 & 38/24 & -8.2/-11 & 38/29 & -13/-17 & 10/5.8 & 11/-11 & 66/46 & -10/-12 & 13/8.2\\
 All green (6) & -0.59/-2.4 & 0.0/0.0 & 18/19 & 6.0/3.9 & 76/8.3 & 0.53/18 & -4.3/-4.0 & 1.2/0.80 & 83/94 & 0.0/0.0 & -4.3/-4.0 & 1.2/0.80\\
 Yellow and green MAD (12) & 0.67/-0.094 & 0.0/0.0 & 21/21 & 4.1/3.0 & 32/0.38 & 16/21 & -5.7/-5.3 & 2.5/1.7 & 63/41 & 11/27 & 3.0/3.9 & 4.4/2.2\\
 Yellow and green SANE (23) & 1.3/1.7 & 0.0/0.37 & -0.21/-15 & 25/24 & -22/-11 & 32/29 & -17/-17 & 8.6/5.4 & -25/-54 & 30/31 & -12/-14 & 11/7.2\\
\enddata
\end{deluxetable*}
\end{rotatetable}
\end{table}
\clearpage
\section{Calibration Set}\label{sec: Calibration Set}
The RIAF-trained \alinet can accurately recover the images and parameters of other RIAFs.  However, the performance on GRMHD simulations result in anticipated ambiguities (multimodal posteriors) and increased systematic uncertainties associated with the different underlying models of the emission regions.  Here we apply the \alinet model to large set of GRMHD simulations \citep[the calibration set from][]{sgrA_paper4} to quantify this systematic uncertainty, thereby generating an effective calibration of the RIAF model library.  
Because we do not necessarily have any a priori reason to trust the GRMHD simulation library more than the RIAF library, this calibration is necessarily dependent on the priors assigned to the various model classes.  Thus, in practice we provide a set of calibrations.

This process is complicated by the multimodal nature of the posteriors and the two distinct RIAF model sets employed (optically thick and thin).  Therefore, we begin with an explicit discussion of how we assess posterior consistency and how we use this to ultimately infer an additional systematic uncertainty to add to the parameter estimates.

We perform the calibration for each parameter independently.  We start by approximating the desired posterior parameter distribution from \themis for some parameter $x$, marginalized over all other parameters, as Gaussian mixture models (GMMs), which are distributions composed of multiple Gaussian components. A typical parameter distribution from \themis can be expressed as:
\begin{equation}
    q_{i}(x) = \sum_{j = 1}^n \frac{w_{i, j}}{\sqrt{2\pi \sigma_{i, j}^2})} \exp{\left[- \frac{(x - \bar{x}_{i, j})^2}{2 \sigma_{i, j}^2}\right]}, 
\end{equation}
where $n$ denotes the total number of Gaussian components in the GMM, $j$ indexes each Gaussian component, and $i$ identifies the specific calibration dataset that $q_i(x)$ represents. In the GRMHD calibration set, there are 90 simulations, so $i \in [0, 89]$. Each Gaussian component has its own mean $\bar{x}_{i, j}$ and standard deviation $\sigma_{i, j}$, with $w_{i, j}$ representing the weight of each component, indicating its contribution to the overall posterior probability distribution.

To introduce the additional parameter-specific systematic error, we convolve each parameter-specific GMM with another Gaussian distribution, $\mathcal{N}(\bar{x}, \sigma)$, the parameters of which have yet to be determined at this stage,  i.e.,

\begin{multline} 
    q_i(x; \bar{x}, \sigma) \equiv q_i (x) * \mathcal{N}(\bar{x}, \sigma) =\\
    \sum_{j=1}^n \frac{w_{i, j}}{\sqrt{2 \pi (\sigma_{i, j}^2 + \sigma^2)}} \exp{\left[-\frac{(x - \bar{x}_{i, j} - \bar{x})^2}{2 (\sigma_{i, j}^2 + \sigma^2)}\right]}. 
\end{multline}
The calibration is performed by determining the $\bar{x}$ and $\sigma$ that result in confidence levels that match those found via the analysis of the calibration set.  Because the tails of the posteriors are not necessarily Gaussian, we repeat this procedure to obtain 2$\sigma$ and 3$\sigma$ calibrations, obtained by matching the $95.45\%$ and $99.73\%$ confidence levels.

Explicitly, this is done by solving for the posterior probability density, $q_{\gamma,i}$, associated with the desired confidence level, $\gamma$, i.e.,
\begin{equation}
    \int_{-\infty}^{+\infty} dx \,\, q_i(x; \bar{x}, \sigma) \,\, \Theta [q_i(x) - q_{\gamma, i}] = \gamma, ~\forall~ i \label{eq: conf interval}
\end{equation}
 where $\Theta (x)$ is the Heaviside step function.
We then obtain $\bar{x}$ and $\sigma$ by requiring that the posterior weight above $q_{\gamma,i}$ matches confidence level of the calibration set, i.e., solving 
\begin{equation} 
    \sum_{i = 0}^{m - 1} \Theta (q_i(x_{{\rm truth}, i};\bar{x},\sigma) - q_{\gamma, i}) = \ceil{\gamma m}, \label{eq: conf level all points}
\end{equation}
where $m=90$ is the number of calibration data sets. That is, our objective is to find $\bar{x}$ and $\sigma$ such that \autoref{eq: conf interval} and \autoref{eq: conf level all points} hold true simultaneously (i.e., a fraction $\gamma$ of the time, a fraction $\gamma$ of the truth values from the calibration sets are within the $\gamma$ confidence level.)
We solve numerically for all $q_{\gamma, i}$, $\bar{x}$, and $\sigma$. We start with an initial guess for $\bar{x}$ and $\sigma$, and find the values for all $q_{\gamma, i}$ such that \autoref{eq: conf interval} is satisfied. We then use the values found for $q_{\gamma, i}$ to approximate the values for $\bar{x}$ and $\sigma$ via differential evolution algorithm \citep{diff_evolution} to satisfy \autoref{eq: conf level all points}. We repeat these two steps until convergence, i.e., until we find values for $q_{\gamma, i}$'s and $\bar{x}$ and $\sigma$ such that both \autoref{eq: conf interval} and \autoref{eq: conf level all points} are satisfied together. This $\bar{x}$ and $\sigma$ denote the added uncertainty due to scattering and model misspecification within confidence level $\gamma$. 

The uncertainty values of different ways of slicing the calibration set of \cite{sgrA_paper4} based on how many of the constraints posed by EHT observations of Sgr A* are satisfied (pizza plots in figure 27 of \cite{sgrA_paper5}), are reported in \autoref{table: Calibration set combined} when a combination of an optically thick and thin RIAF are used for fitting; all values listed should be multiplied by $10^{-2}$.

The parameters for which calibration is done are flux intensity ratio, mass ratio, spin magnitude, spin axis, spin value, and normalized spin vector. Note that here, we calibrate for normalized spin vector and spin axis instead of position angle (PA) and cosine of inclination angle ($\mu$)\footnote{Note that $\mu$ refers not to an average value but to cosine of an angle.} due to the inherent degeneracies that exist in defining the orientation of the spin vector with inclination angle and position angle; for example, PA angle is $2\pi$ periodic. Furthermore, different combinations of inclination angle and position angle might lead to the same spin vector orientation. Since the more physically important parameter of interest is the normalized spin vector itself, we do direct calibration on this parameter. For a more exhaustive list of uncertainties of different slices of the calibration set, as well as the uncertainty introduced only from scattering and/or from fitting with a generative model to RIAF simulations, look at \autoref{sec: Calibration appdx}. 

Note that, due to the limitation in the number of data points in the calibration set, some of the confidence intervals are not very accurate. For example, the number of data points in the calibration set that pass all the EHT constraints (the green in figure 27 of \cite{sgrA_paper4}) is only $6$ in total, well below what is needed to reliably define a $2\sigma$ confidence interval, which is about $20$ data points.

\section{Summary and Conclusions}\label{sec: conclusions}
This study aimed to explore the robustness and limitations of semi-analytical models, specifically radiatively inefficient accretion flows (RIAFs), in fitting and interpreting Event Horizon Telescope (EHT) observations. Through the development and application of \alinet, a generative machine learning model, we have demonstrated significant advancements in efficiently generating synthetic images and constraining the physical parameters of black hole accretion flows. 

Our self-consistency tests confirmed that the \alinet model, when trained on RIAF data, could accurately reproduce the physical parameters of synthetic images generated by the model itself. This establishes \alinet as a reliable tool for parameter inference within the parameter space spanned by the RIAF model. The success of this approach underscores the viability of employing machine learning techniques in astrophysical data analysis, particularly in contexts where traditional methods are computationally prohibitive.

We further extended our analysis by evaluating the impact of scattering effects on the ability of the RIAF-trained \alinet to recover physical parameters. The results showed that while scattering introduces additional uncertainty, the combination of \alinet and \themis remains capable of effectively constraining the physical properties of the system, which, as previously known and implemented within the EHT, illustrates the importance of accounting for scattering in the interpretation of EHT data, where such effects are intrinsic to the observed signals.

A key aspect of our study was the investigation of robustness to model misspecification, particularly the comparison between the static RIAF model and variable GRMHD simulations. By applying the RIAF-trained \alinet to GRMHD data, we were able to assess the degree of compatibility between these two models. Our results indicate that, while RIAFs can provide a reasonable approximation for certain configurations, significant discrepancies arise in cases involving more dynamic and jet-dominated flows. These discrepancies highlight the limitations of using stationary and disk-dominated models such as static RIAFs for interpreting dynamic or jet-dominated observational data and underscore the importance of carefully assessing the applicability of approximate models in complex astrophysical environments. Future work could explore ways to refine these models and incorporate more sophisticated simulations to improve the accuracy and reliability of the interpretation of EHT data.

With calibration of \alinet-generated and \themis-fitted RIAFs on a variety of datasets, we quantified the prediction uncertainty under different assumptions. For example, assuming RIAFs are the underlying physical model explaining the data, the prediction uncertainty, e.g., for the value of spin is about $10\%$ within $2\sigma$ confidence interval. However, if we assume that the uncerlying physical model is a MAD or SANE GRMHD, then the $2\sigma$ confidence interval uncertainty rises to about $60\%$ for the spin value. The relevant uncertainties are required to be included in any and all future analyses of the physical parameters of the EHT observational data (look at \autoref{table: Calibration set combined} for all the confidence levels).

The next step in this study is to apply this framework to observational data of Sgr A* from the 2017 EHT observations \citep{sgrA_paper1}. By leveraging the calibrated uncertainties obtained in this work, we aim to constrain the physical parameters of Sgr A* while accounting for the systematic effects identified in our robustness tests. The combination of \alinet, \themis, and our calibration procedure will allow us to provide well-motivated parameter estimates with appropriately quantified uncertainties. This approach will be instrumental in refining our understanding of the accretion flow properties and the fundamental physics governing the event-horizon-scale environment of Sgr A*.

\begin{acknowledgements}
We thank Ben Prather, Abhishek Joshi, Vedant Dhruv, C.K. Chan, and Charles Gammie for the synthetic images used here, generated under NSF grant AST 20-34306.  This research used resources of the Oak Ridge Leadership Computing Facility at the Oak Ridge National Laboratory, which is supported by the Office of Science of the U.S.\ Department of Energy under Contract No. DE-AC05-00OR22725. This research used resources of the Argonne Leadership Computing Facility, which is a DOE Office of Science User Facility supported under Contract DE-AC02-06CH11357. This research was done using services provided by the OSG Consortium, which is supported by the National Science Foundation awards \#2030508 and \#1836650. This research is part of the Delta research computing project, which is supported by the National Science Foundation (award OCI 2005572), and the State of Illinois. Delta is a joint effort of the University of Illinois at Urbana-Champaign and its National Center for Supercomputing Applications.
The authors thank Boris Georgiev and Rufus Ni for helpful comments. This work was supported in part by Perimeter Institute for Theoretical Physics.  Research at Perimeter Institute is supported by the Government of Canada through the Department of Innovation, Science, and Economic Development Canada and by the Province of Ontario through the Ministry of Economic Development, Job Creation and Trade.
A.E.B. receives additional financial support from the Natural Sciences and Engineering Research Council of Canada through a Discovery Grant.
This research was enabled in part by the support provided by Calcul Qu'ebec (www.calculquebec.ca) and the Digital Research Alliance of Canada (alliancecan.ca).
\end{acknowledgements}
\clearpage
\bibliography{main}{}
\bibliographystyle{aasjournal}

\appendix 

\section{Comprehensive Library of Tests}\label{sec: Tests appdx}
\begin{figure*}[ht!]
    \centering
    \fig{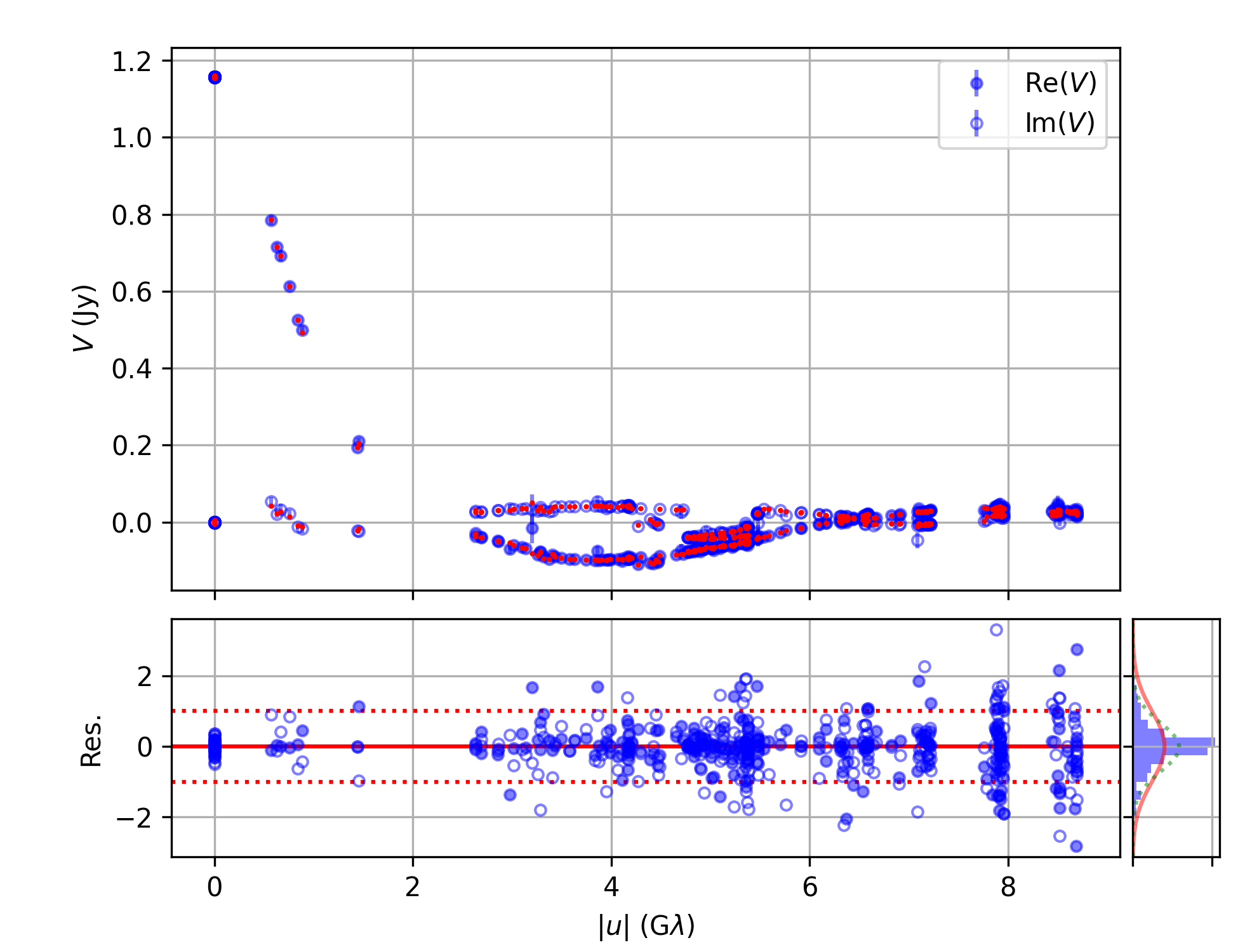}{0.46\textwidth}{(a) Self Test.}
    ~
    \fig{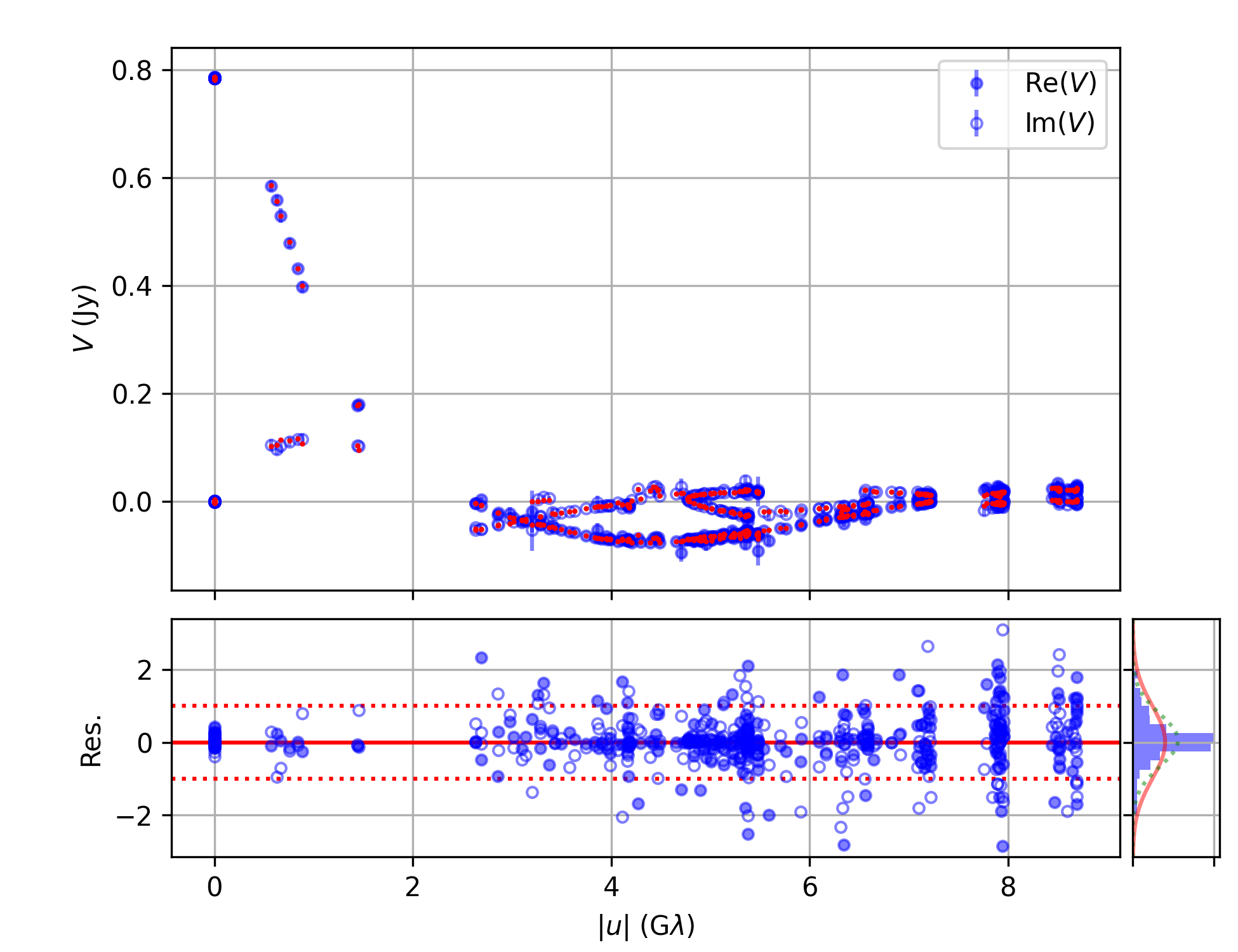}{0.45\textwidth}{(b) RIAF Test.}
    
    \fig{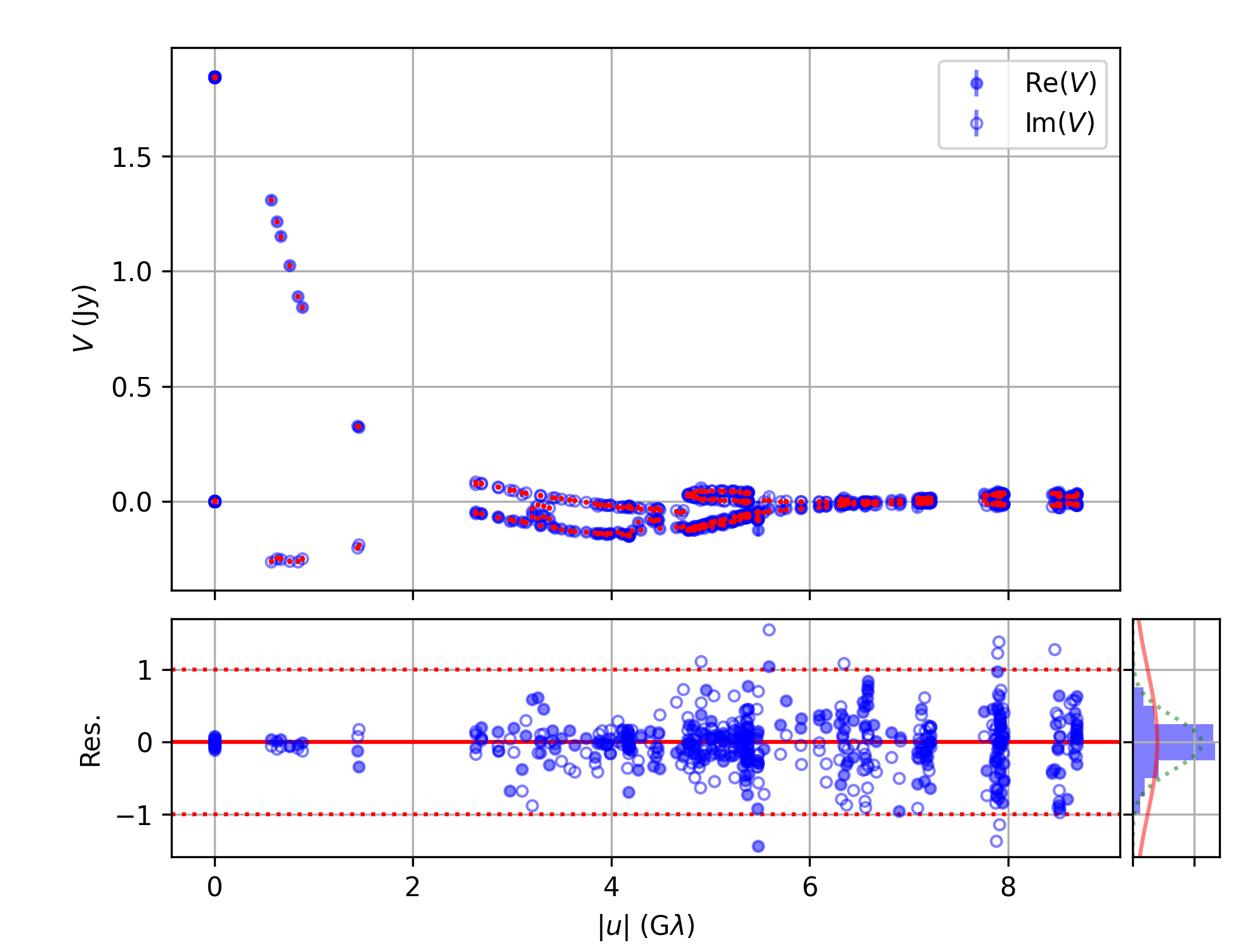}{0.45\textwidth}{(c) Scattering RIAF Test with threshold noise of $7$mJy.}
    ~
    \fig{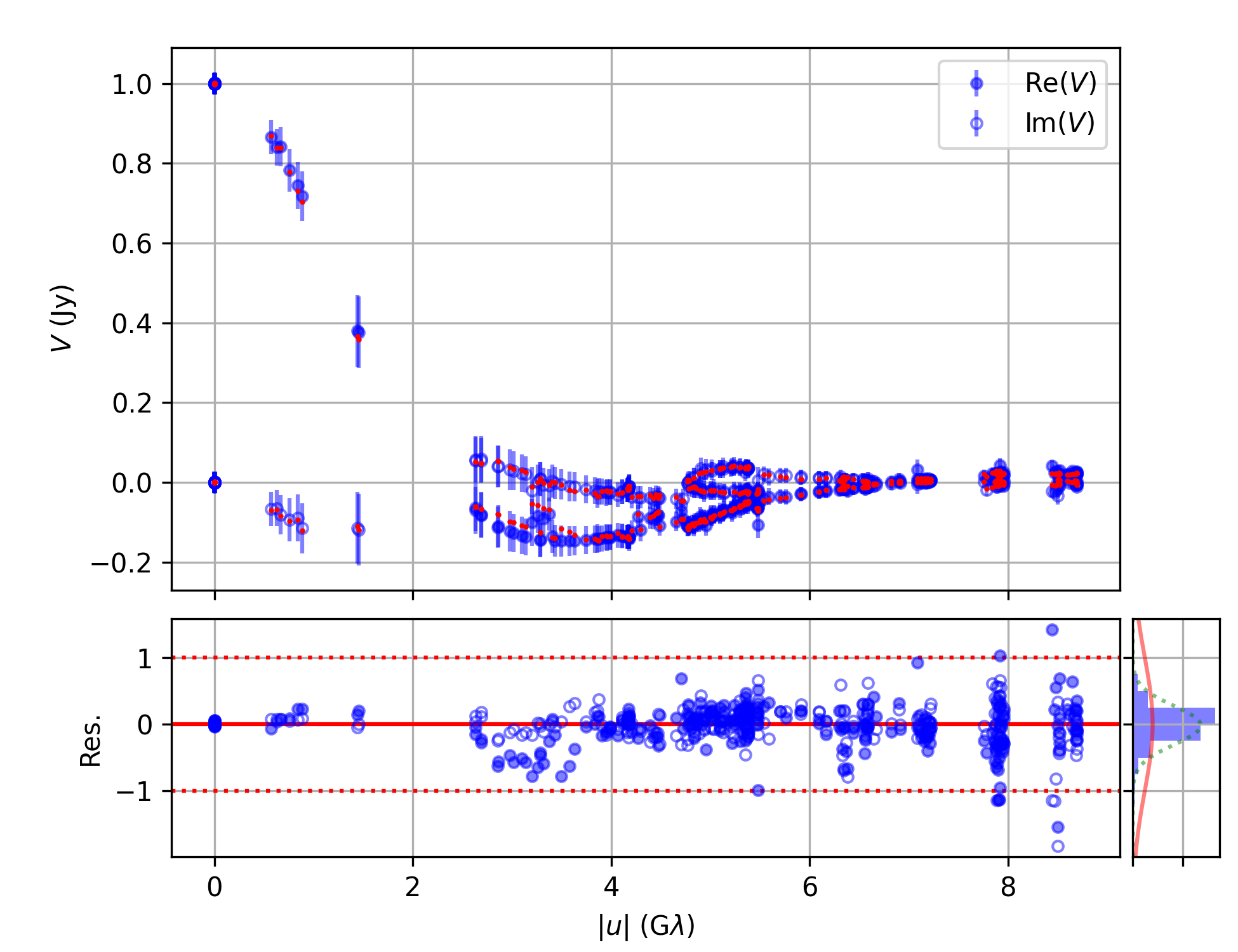}{0.45\textwidth}{(d) Mean GRMHD Validation Set with threshold noise of $20$mJy.}
    \caption{Top section: Each panel presents the normalized complex visibility data (blue) as a function of baseline length, denoted as $V$. The real and imaginary components of the complex visibilities are shown as filled and open markers, respectively, with the corresponding model visibilities overlaid in red. These data have undergone pre-analysis and pre-imaging calibration as described in \citet{m87_paper3, m87_paper2, m87_paper1}. Bottom section: Each panel depicts the normalized residuals, computed as the difference between the model and observed visibilities, divided by the observational uncertainties, and plotted against baseline length. A solid red horizontal line marks zero residual, while the two dotted red horizontal lines indicate $\pm 1$ standard deviation. To the right of each bottom panel, a blue histogram visualizes the distribution of the normalized residuals, with a solid red curve representing a unit-variance normal distribution and a dotted green curve corresponding to a normal distribution with variance matching that of the residuals. These residual maps serve as a diagnostic tool for evaluating how well \alinet within \themis recovers visibility data. Complex visibility consists of both amplitude and phase, corresponding to a specific baseline in the interferometer array. The amplitude encodes information about the contrast between different regions of the source, while the phase conveys positional information relative to the baseline orientation. As demonstrated here, the ability of \alinet within \themis to reconstruct both the amplitude and phase of the complex visibilities highlights its effectiveness in capturing the spatial structure of the source.\label{fig: RIAF visibilities}}
\end{figure*}
\subsection{Complex Visibility Residuals}\label{subsec: Complex visibility residuals appdx}
\autoref{fig: RIAF visibilities} shows the complex visibility residual maps of the representative test case discussed for each test in the main text, namely, \alinet self test, RIAF test, scattered RIAF test, and scattered mean GRMHD model robustness test. Complex visibility is a complex number comprising both amplitude and phase, corresponding to a specific baseline (the distance and orientation between two telescopes) in the interferometer array. The amplitude indicates the strength of the signal at a particular spatial frequency, reflecting the contrast between different regions of the source, while the phase provides information about the position of structures within the source relative to the baseline. \autoref{fig: RIAF visibilities} demonstrates the ability of \alinet within \themis to recover the information encoded in the amplitude and phase of the complex visibilities. 

\begin{figure*}[ht!]
    \centering
    \fig{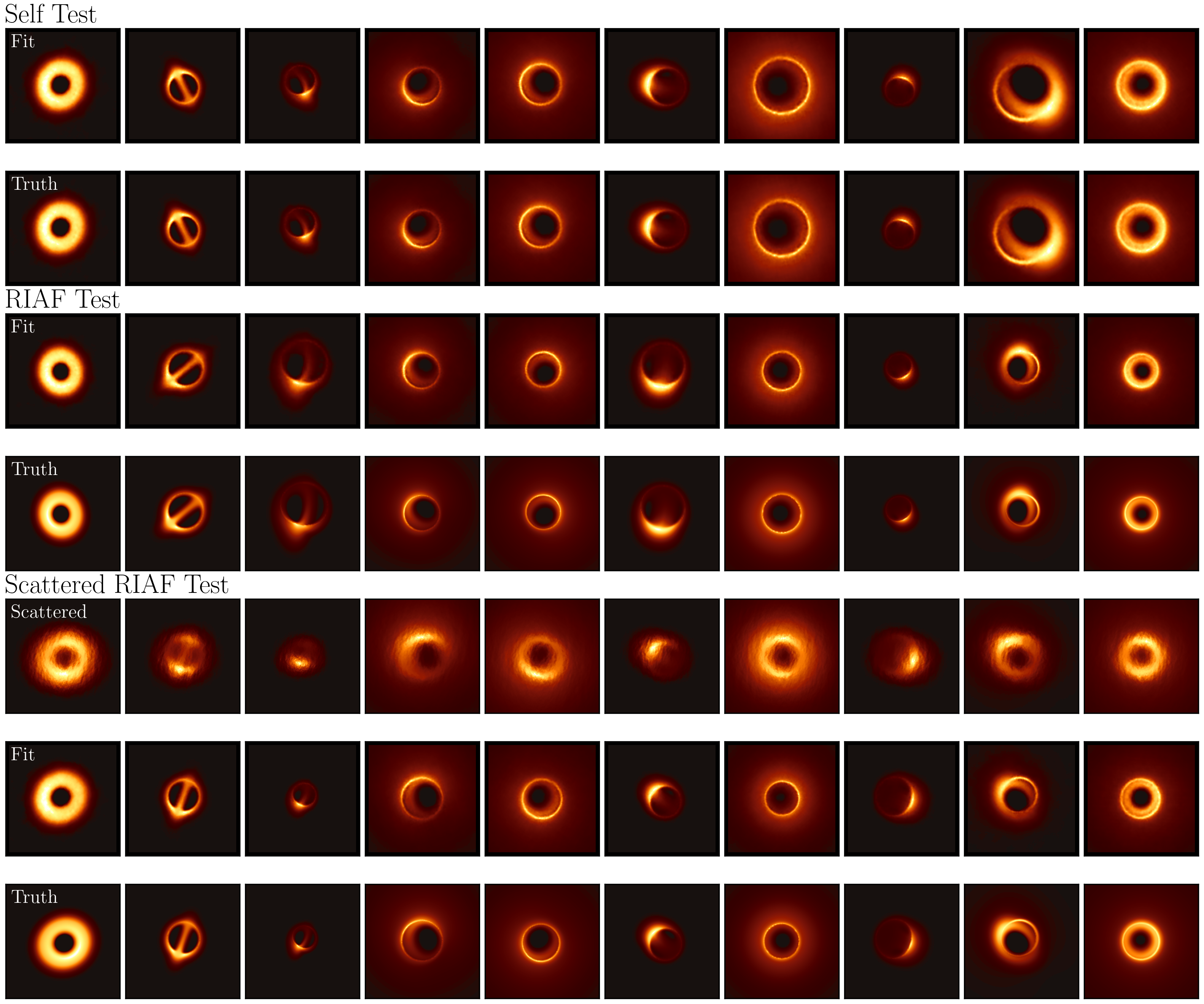}{\textwidth}{}
    \caption{Fitting results from the \alinet self test, RIAF test, and scattered RIAF test. A threshold noise level of $7$ mJy is applied in the scattered test. The rows, from top to bottom, present: the fitting results for the \alinet self test, their corresponding ground truths, the fitting results for the RIAF test, the ground truth RIAF images, the fitting results for the scattered RIAF test, the recovered scattered RIAF images, and their unscattered ground truth counterparts. These results demonstrate that \alinet within \themis effectively recovers the underlying image across a range of physical parameters.\label{fig: RIAF tests fit_and_truth}},
\end{figure*}

\begin{figure*}[hb!]
    \centering
    \fig{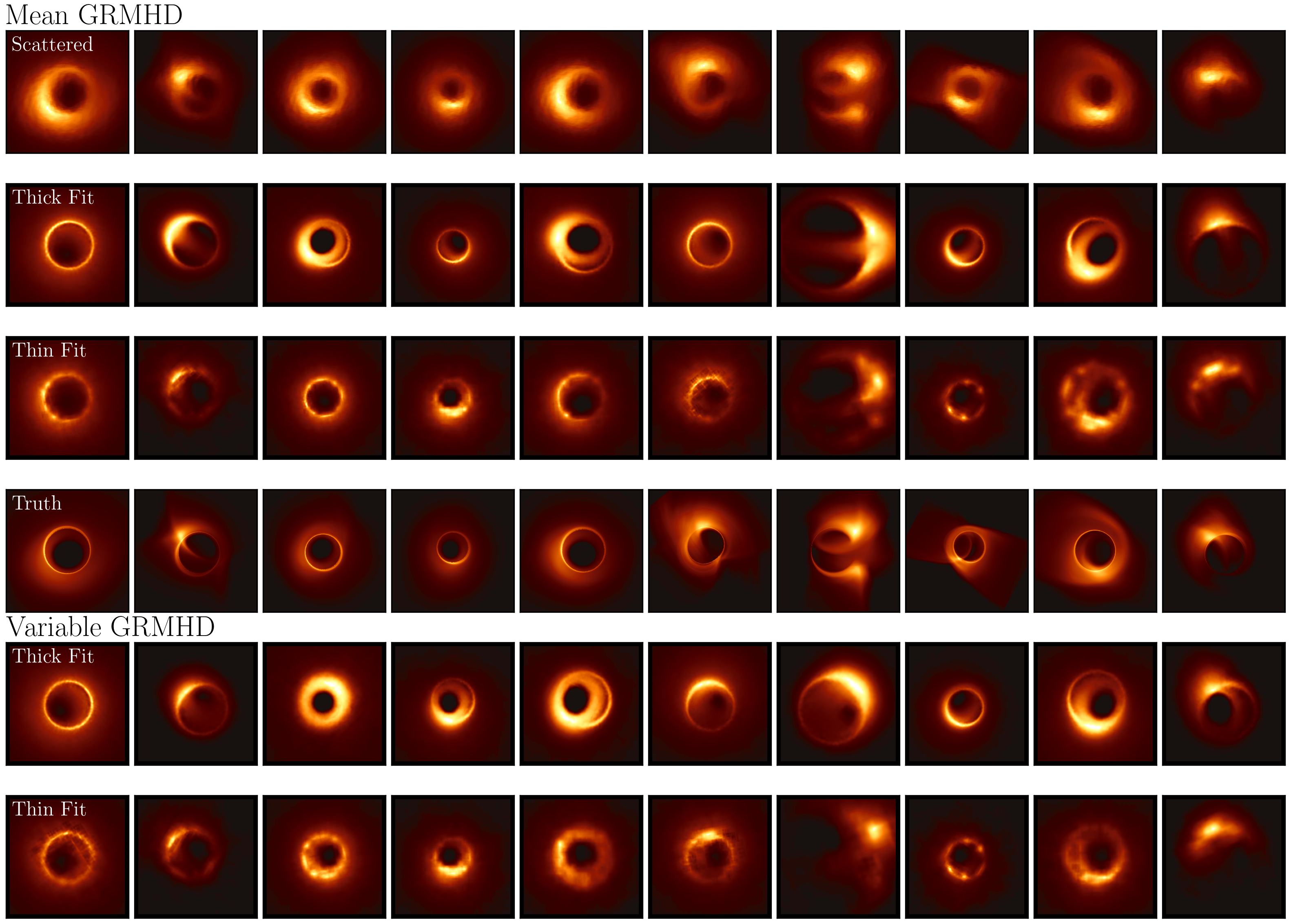}{\textwidth}{}
    \caption{Fitting results for scattered mean GRMHD and scattered variable GRMHD validation set tests, using a $20$mJy threshold for noise. The first row shows the scattered ground truth mean GRMHD images from the validation set. The second and third rows display the results of fitting optically thick and optically thin RIAF models to the scattered mean GRMHD simulations. The fourth row presents the unscattered mean GRMHD ground truth. The last two rows correspond to optically thick and thin RIAF fitting results applied to the scattered variable GRMHD validation set. Since the variable case represents a sequence of frames rather than a single snapshot, there is no direct ground truth; instead, the fourth row provides the mean image, with its scattered counterpart shown in the first row.\label{fig: val GRMHD tests fit_and_truth}}
\end{figure*}

\begin{figure*}[h!]
    \centering
    \includegraphics[width=0.8\linewidth]{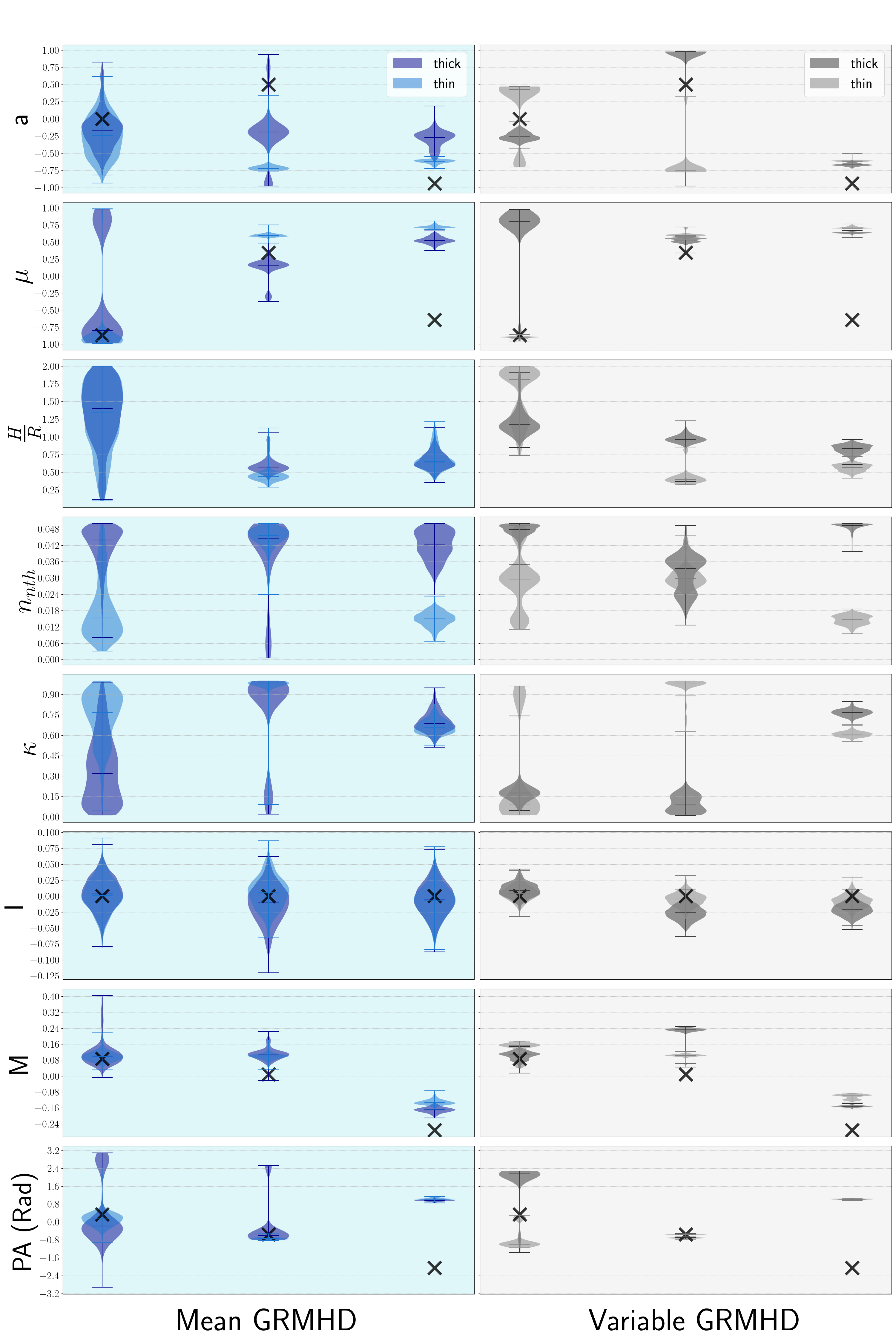}
    \caption{Violin plots for three test cases for mean GRMHD (middle) and variable GRMHD (right). The X shows the truth value. The violins show the resulting fitted distributions from \themis. The bars in each distribution show minimum, maximum, and the median values of the distribution. In each column, the first two from the left are MADs, and the third is a SANE. Distributions for thick and thin RIAFs fitted to GRMHDs are shown separately.}
    \label{fig: grmhd violins appdx}
\end{figure*}

\subsection{RIAF Tests}\label{subsec: RIAF Tests appdx}
The fitting process of \alinet within \themis for \alinet self tests and RIAF tests involves 1. generating simulated observational data, 2. using \themis to generate data from the images output from \alinet, 3. Finding the posteriors over the images and parameters of all the simulated data generated and compared with the truth data point. For scattered RIAF test, the process is very similar, with the caveat of an added layer of scattering mitigation, where the comparison is done between the (scattered) truth RIAF and the scattered data generated from \alinet within \themis. \autoref{fig: RIAF tests fit_and_truth} shows the truth images along with their best fit found from the posteriors for the self test, RIAF test, and scattered RIAF test. Note that even with the addition of scattering error, \alinet within \themis is still able to recover the underlying truth image, which is most often exactly the peak of the posteriors. 
Moreover, \autoref{fig: grmhd violins appdx} shows the posteriors of fitting optically thick and thin RIAFs separately to the mean and variable GRMHDs discussed in \autoref{fig: scatt,grmhd violins}. Therefore, the source of multimodality is multiple, one is that each of the posteriors is multimodal on its own due to scattering and model misspecification, and second is that the concatenation of two separate posteriors leads to more multimodality. We see in these test cases that putting the posteriors of both models together leads to more certainty of the fits, which is further evidence for the need of a more comprehensive RIAF simulation which incorporates a clumping factor. 
\subsection{GRMHD Tests}\label{subsec: GRMHD Tests appdx}
The fitting process of \alinet within \themis for the GRMHD tests is similar to the scattered RIAF tests, with the only difference being that this time, RIAF-trained \alinet models are being used to recover the underlying phyiscs of GRMHD simulated data. \autoref{fig: val GRMHD tests fit_and_truth} visualizes the result of fitting both the optically thick and thin RIAF-trained \alinet to the average GRMHD images of the validation set of \citet{sgrA_paper4}. Even though the RIAF-trained models are doing their best to create similar looking images to the GRMHDs, the inherent differences in assumptions in RIAFs and GRMHDs leads to some of the RIAF images looking very different from the GRMHDs to which fitting is being done.

\begin{table}[p]
\vspace{5cm}
\begin{rotatetable}
    \begin{deluxetable*}{lcc|cc|cc|cc|cc|ccr}
\tabletypesize{\scriptsize}
\tablewidth{0pt}
\tablecaption{A comprehensive list of model misspecification and scattering uncertainties on the variable GRMHD calibration set for thick and thin RIAF models, separately as well as together. The number in the parentheses in front of each dataset category denotes the number of samples that were used to quantify the uncertainties. \label{table: Calibration set combined appdx}}
\tablehead{
 \colhead{Dataset} & \multicolumn{2}{c}{Flux $I$} & \multicolumn{2}{c}{Mass $M$} & \multicolumn{2}{c}{Spin magnitude $|a|$} & \multicolumn{2}{c}{Spin axis $z_a$} & \multicolumn{2}{c}{Spin value $a$} & \multicolumn{2}{c}{Normalized spin vector $n_a$}\\
  \colhead{$(\times 1e-2)$} & \colhead{$\Delta_I$} & \colhead{$\sigma_I$} & \colhead{$\Delta_M$} & \colhead{$\sigma_M$} & \colhead{$\Delta_a$} & \colhead{$\sigma_a$} & \colhead{$\Delta_{z_a}$} & \colhead{$\sigma_{z_a}$} & \colhead{$\Delta_{a}$} & \colhead{$\sigma_{a}$} & \colhead{$\Delta_{n_a}$} & \colhead{$\sigma_{n_a}$} & \\
  & \colhead{$2\sigma$/$3\sigma$} & \colhead{$2\sigma$/$3\sigma$} & \colhead{$2\sigma$/$3\sigma$} & \colhead{$2\sigma$/$3\sigma$} & \colhead{$2\sigma$/$3\sigma$} & \colhead{$2\sigma$/$3\sigma$} & \colhead{$2\sigma$/$3\sigma$} & \colhead{$2\sigma$/$3\sigma$} & \colhead{$2\sigma$/$3\sigma$} & \colhead{$2\sigma$/$3\sigma$} & \colhead{$2\sigma$/$3\sigma$} & \colhead{$2\sigma$/$3\sigma$}
}
\startdata 
\multicolumn{14}{l}{\textbf{Thick RIAF}}\\
\hline
Thick RIAF (10) & -0.97/ & 0.0/ & -0.58/ & 0.55/ & -5.3/ & 1.3/ & -0.26/ & 0.0/ & -2.8/ & 3.5/ & 0.19/ & 0.0/\\
Scattered RIAF (10) & 4.0/ & 0.0/ & 0.90/ & 0.0/ & 9.0/ & 9.0/ & -0.25/ & 0.0/ & -8.5/ & 10/ & -19/ & 15/\\
 Entire calibration set (90) & 0.30/1.7 & 0.41/1.1 & -18/36 & 30/39 & -6.7/-2.4 & 44/37 & -40/-48 & 21/17 & -5.9/-23 & 71/75 & -29/-2.2 & 29/33\\
 All yellow and green (35) & 0.62/1.9 & 0.0/0.96 & -22/-12 & 30/24 & -7.1/-9.5 & 43/33 & -46/-48 & 24/17 & 23/-24 & 60/57 & -26/-2.2 & 36/33\\
 All green (6) & 0.18/0.67 & 0.0/0.0 & 11/-11 & 11/16 & 64/-5.4 & 0.0/28 & -4.1/-3.7 & 1.4/1.1 & -63/21 & 9.5/36 & -4.1/-3.7 & 1.4/1.1\\
 All MAD (45) & 0.24/1.9 & 0.0/0.96 & 11/17 & 18/12 & 7.3/-2.4 & 43/37 & -37/-43 & 19/15 & 7.0/-0.28 & 65/49 & -24/-29 & 27/21\\
 Yellow and green MAD (12) & 0.17/0.27 & 0.0/0.0 & 8.4/8.4 & 13/8.2 & 17/0.59 & 19/29 & -13/-43 & 6.5/15 & 65/27 & 34/42 & 10/-29 & 8.3/21\\
 Green MAD (4) & 0.92/1.1 & 0.0/0.0 & 11/11 & 13/7.2 & 64/64 & 0.0/0.0 & -0.72/-0.38 & 0.0/0.0 & 100/100 & 12/5.8 & -0.64/-0.30 & 0.0/0.0\\
 All SANE (45) & 0.30/1.7 & 0.41/1.1 & -27/6.1 & 34/32 & -12/-7.0 & 45/34 & -40/-48 & 21/17 & -1.1/-26 & 72/74 & -16/-23 & 33/38\\
 Yellow and green SANE (23) & 0.59/1.9 & 0.0/0.96 & -23/-12 & 34/24 & -20/-18 & 39/29 & -47/-49 & 23/17 & -1.1/-24 & 76/57 & -26/-2.2 & 37/33\\
 Green SANE (2) & 0.24/-2.3 & 0.0/0.0 & -21/-30 & 9.0/8.4 & -63/-64 & 10/6.5 & -8.4/-9.6 & 0.0/0.0 & -64/-64 & 10/6.5 & -8.0/-9.6 & 0.0/0.0\\
\hline
\multicolumn{14}{l}{\textbf{Thin RIAF}}\\
\hline
Thin RIAF (10) & 1.5/ & 1.4/ & 8.9/ & 4.7/ & 3.0/ & 12/ & -2.9/ & 1.1/ & 11/ & 9.1/ & -2.8/ & 1.0/\\
Scattered RIAF (10) & -0.23/ & 0.4/ & -0.8/ & 16/ & 15/ & 22/ & -1.3/ & 0.4/ & 10/ & 27/ & 0.6/ & 0.7/\\
Entire calibration set (90) & 0.28/1.0 & 0.42/1.0 & -1.9/-19 & 24/25 & 7.5/1.8 & 40/33 & -38/-49 & 21/18 & -33/-9.3 & 67/63 & -7.4/-4.4 & 37/36\\
 All yellow and green (35) & -0.50/0.93 & 0.0/0.70 & -8.4/-19 & 29/25 & 6.1/1.8 & 41/32 & -40/-50 & 28/18 & 14/-14 & 54/45 & -24/-9.3 & 45/33\\
 All green (6) & -2.0/-2.0 & 0.0/0.0 & 19/19 & 5.6/3.6 & 6.9/82 & 4.8/0.0 & -4.2/-4.2 & 1.3/0.73 & 81/37 & 0.0/27 & -4.2/-4.2 & 1.3/0.72\\
 All MAD (45) & 1.1/1.4 & 0.0/0.92 & 12/9.4 & 9.4/7.0 & -5.8/4.7 & 39/32 & -37/-47 & 21/17 & -32/-8.8 & 64/48 & 24/-4.4 & 34/36\\
 Yellow and green MAD (12) & -0.22/-1.0 & 0.0/0.0 & 12/14 & 9.4/5.5 & 50/16 & 18/26 & -5.9/-39 & 1.9/13 & 65/40 & 26/25 & 0.92/1.6 & 4.7/3.1\\
 Green MAD (4) & -1.8/0.19 & 0.0/0.0 & 24/27 & 2.0/0.37 & 80/83 & 2.9/0.0 & -4.2/-4.2 & 1.3/0.74 & 81/99 & 0.0/4.2 & -4.2/-4.2 & 1.3/0.74\\
 All SANE (45) & -0.35/0.64 & 0.0/0.87 & -14/-19 & 26/25 & 3.7/1.1 & 40/33 & -40/-49 & 21/18 & 34/-16 & 62/59 & -7.4/-9.3 & 37/33\\
 Yellow and green SANE (23) & 0.66/1.2 & 0.0/0.52 & -8.3/-19 & 29/25 & 3.2/-1.1 & 42/32 & -47/-50 & 25/18 & 14/-14 & 55/45 & 16/-9.3 & 40/33\\
 Green SANE (2) & -3.3/1.7 & 0.0/0.0 & 1.2/13 & 0.0/0.0 & -33/-33 & 2.5/1.8 & -6.8/-6.9 & 0.24/0.23 & -33/-33 & 2.5/1.8 & 95/-33 & 0.0/0.0\\
 \hline
 \multicolumn{14}{l}{\textbf{Thick + Thin RIAF}}\\
 \hline 
 Entire calibration set (90) & -0.74/1.9 & 0.0/0.80 & -12/-15 & 20/24 & -4.0/0.86 & 39/34 & -28/-43 & 17/15 & -28/-27 & 59/61 & -13/-29 & 21/21\\
 All yellow and green (35) & 1.3/1.5 & 0.0/0.52 & -1.1/-15 & 38/24 & -8.2/-11 & 38/29 & -13/-17 & 10/5.8 & 11/-11 & 66/46 & -10/-12 & 13/8.2\\
 All green (6) & -0.59/-2.4 & 0.0/0.0 & 18/19 & 6.0/3.9 & 76/8.3 & 0.53/18 & -4.3/-4.0 & 1.2/0.80 & 83/94 & 0.0/0.0 & -4.3/-4.0 & 1.2/0.80\\
 All MAD (45) & -0.53/1.9 & 0.0/0.79 & 15/9.6 & 7.4/7.1 & -0.095/10 & 32/30 & -11/-43 & 9.1/15 & 4.8/13 & 60/52 & 2.2/-29 & 14/21\\
 Yellow and green MAD (12) & 0.67/-0.094 & 0.0/0.0 & 21/21 & 4.1/3.0 & 32/0.38 & 16/21 & -5.7/-5.3 & 2.5/1.7 & 63/41 & 11/27 & 3.0/3.9 & 4.4/2.2\\
 Green MAD (4) & 0.32/-1.7 & 0.0/0.0 & 33/33 & 0.0/0.0 & 76/81 & 0.53/0.0 & -1.3/-0.39 & 0.0/0.0 & 100/94 & 8.3/0.0 & -1.3/-0.43 & 0.0/0.0\\
 All SANE (45) & -1.6/1.6 & 0.0/0.42 & -12/-15 & 31/24 & -13/-11 & 34/29 & -26/-32 & 15/11 & -27/-14 & 50/45 & -22/-19 & 19/17\\
 Yellow and green SANE (23) & 1.3/1.7 & 0.0/0.37 & -0.21/-15 & 25/24 & -22/-11 & 32/29 & -17/-17 & 8.6/5.4 & -25/-54 & 30/31 & -12/-14 & 11/7.2\\
 Green SANE (2) & 1.2/-5.0 & 0.0/0.0 & 8.3/13 & 0.43/0.0 & -40/-38 & 0.0/0.0 & -8.7/-9.2 & 0.0/0.0 & -38/-38 & 0.0/0.0 & -6.0/-6.4 & 0.0/0.0\\
\enddata
\end{deluxetable*}
\end{rotatetable}
\end{table}

\section{Calibration Data Uncertainty Quantification}\label{sec: Calibration appdx}
\autoref{table: Calibration set combined appdx} is a more extensive version of \autoref{table: Calibration set combined}. \autoref{table: Calibration set combined appdx} delineates the additional $2\sigma$ and $3\sigma$ uncertainties of fitting both optically thick and thin RIAF models separately and together to RIAFs and GRMHDs. The datasets for which the uncertainty of fitting the optically thick and thin RIAFs are calculated are RIAFs themselves, scattered RIAFs, the entire calibration set of \citet{sgrA_paper4}, as well as all the yellow and green cases, the green cases, all MAD cases, yellow and green MAD cases, green MAD cases, all SANE cases, yellow and green SANE cases, and green SANE cases. The green and yellow here are a way to categorize the data on how many of the EHT constraints they pass (or fail), where green means they pass all constraints, and yellow means they pass most of the constraints (look at \citet{sgrA_paper5} for details). Note that for the uncertainty quantification of RIAFs, we only mention the $2\sigma$ confidence interval, since the errors are small and a $3\sigma$ lead to a zero additional uncertainty. Furthermore, it is important to note that some of the slicing of the data only leaves a very small subset which is not enough for a meaningful uncertainty quantification within $2\sigma$ and $3\sigma$ confidence levels. We have nevertheless include them here for the sake of completeness. 
\end{document}